\documentclass{mn2e}
\usepackage{latexsym}
\usepackage{graphicx}
\usepackage{amssymb}
\usepackage{epsfig}
\newif\ifAMStwofonts
\AMStwofontstrue

\newcommand{\beqn}{\begin{eqnarray*}}
\newcommand{\eeqn}{\end{eqnarray*}}

\def\be{\begin{equation}}
\def\ee{\end{equation}}
\def\beq{\begin{eqnarray}}
\def\eeq{\end{eqnarray}}
\def\nn{\nonumber}
 
\begin{document}

\title[Rotating neutron stars: an invariant comparison]
{Rotating neutron stars: an invariant comparison of approximate and numerical spacetime models}

\author[
Emanuele Berti, Frances White, Asimina Maniopoulou, Marco Bruni]
{Emanuele Berti${}^{1}$, Frances White${}^{2}$, Asimina Maniopoulou${}^{3}$, Marco Bruni${}^{2}$\\
${}^{1}$ McDonnell Center for the Space Sciences, Department of
Physics, Washington University, St. Louis, Missouri 63130, USA.\\
Present address: Groupe de Cosmologie et Gravitation
(GReCO), Institut d'Astrophysique de Paris (CNRS), $98^{bis}$ Boulevard Arago, \\
75014 Paris, France
\\ ${}^{2}$ Institute of Cosmology and Gravitation, Mercantile House,
Hampshire Terrace, University of Portsmouth, Portsmouth PO1 2EG, UK
\\ ${}^{3}$ School of Mathematics, University of Southampton,
Southampton SO17 1BJ, UK }

\maketitle

\begin{abstract}

We compare three different models of rotating neutron star spacetimes:
i) the Hartle-Thorne (1968) slow-rotation approximation, keeping terms
up to second order in the stellar angular velocity; ii) the exact
analytic vacuum solution of Manko \textit{et al.} (2000a, 2000b); and
iii) a numerical solution of the full Einstein equations.  In the
first part of the paper we estimate the limits of validity of the
slow-rotation expansion by computing relative errors in the
spacetime's quadrupole moment $Q$ and in the corotating and
counterrotating radii of Innermost Stable Circular Orbits (ISCOs)
$R_{\pm}$.  We integrate the Hartle-Thorne structure equations for
five representative equations of state. Then we match these models to
numerical solutions of the Einstein equations, imposing the condition
that the gravitational mass and angular momentum of the models be the
same.  We find that the Hartle-Thorne approximation gives very good
predictions for the ISCO radii, with $R_{\pm}$ accurate to better than
$1\%$ even for the fastest millisecond pulsars. At these rotational
rates the accuracy on $Q$ is $\sim 20\%$, and better for longer
periods. In the second part of the paper we focus on the exterior
vacuum spacetimes, comparing the Hartle-Thorne approximation and the
Manko analytic solution to the numerical models using Newman-Penrose
(NP) coordinate-independent quantities. For all three spacetimes we
introduce a physically motivated `quasi-Kinnersley' NP frame. In this
frame we evaluate a quantity, the speciality index $S$, measuring the
deviation of each stellar model from Petrov Type D. Deviations from
speciality on the equatorial plane are smaller than $5 \%$ at star
radii for the faster rotating models, and rapidly decrease for slower
rotation rates and with distance. We find that, at leading order, the
deviation from Type D is proportional to $(Q-Q_{\rm Kerr})$. Our main
conclusion is that the Hartle-Thorne approximation is very reliable
for most astrophysical applications.

\end{abstract}


\section{Introduction}\label{intro}

Pulsars are observed to rotate with periods as small as 1.56 ms
(Backer {\it et al.} 1982, Kulkarni 1995, Chakrabarty {\it et al.}
2003).  Fast rotation is also very important in newly-born neutron
stars, which may undergo secular and dynamical instabilities
(Andersson \& Kokkotas 2001, Kokkotas \& Ruoff 2002). An accurate
modelling of the spacetime of rotating stars is therefore important
both in astrophysical studies and in investigations of sources of
gravitational waves. The aim of this paper is to compare two
relativistic approximate models, the slow rotation expansion of Hartle
\& Thorne (1968, henceforth HT) and the exact vacuum solution by Manko
{\it et al.} (Manko 2000a, 2000b), with a full general-relativistic
numerical one, quantifying the range of validity of the HT expansion
and studying the possible overlap of the two approximations.

The Schwarzschild metric describes either a non-rotating black hole
(BH) or the exterior of a non-rotating star: this is a well-known
consequence of Birkhoff's theorem. In contrast, the rotating Kerr BH
metric describes a rotating star only to linear order in rotational
velocity. At higher orders, the multipole moments of the gravitational
field created by a rapidly rotating compact star differ from those of
a BH.  This difference in the multipolar structure has important
consequences for the observation of the electromagnetic and
gravitational radiation from these objects.  For example, the
detection of gravitational waves emitted by particles in orbit around
BHs and compact stars can be used, in principle, to map the multipolar
structure of the corresponding spacetime and check the validity of the
BH no-hair theorem (Ryan 1995, 1997; Collins \& Hughes 2004).

A relatively simple description of the spacetime of rotating stars was
developed many years ago by Hartle and collaborators (Hartle 1967,
Hartle \& Thorne 1968; see also Chandrasekhar \& Miller 1974) using a
slow rotation expansion, that is, an expansion in the parameter
$\epsilon=\Omega/\Omega^*$. Here $\Omega$ is the star's angular
velocity and $\Omega^*=(M/R^3)^{1/2}$ is a rotational scale set by the
Keplerian frequency of a test particle sitting on the equator of a
non-rotating star with gravitational mass $M$ and radius $R$
(throughout this paper we use geometrical units, $G=c=1$). In the
Newtonian limit, uniformly rotating polytropic stars shed matter at
the equator at a mass-shedding frequency which is roughly given by
$\Omega_K\simeq (2/3)^{3/2}\Omega^*$. This is not true in a general
relativistic framework (see eg. Stergioulas 2003). However $\Omega^*$
does give an order of magnitude estimate of the mass shedding
frequency, so we can roughly expect the HT treatment to break down
when $\epsilon\sim 1$.

The HT method allows a systematic computation of the interior and
exterior structure at successive orders in $\epsilon$. At order
$\epsilon$, deviations from non-rotating stellar models are due to the
relativistic frame dragging, but the stellar structure is still
spherical. Deviations from sphericity are measured by the star's
quadrupole moment; they appear at order $\epsilon^2$.  Being a
slow-rotation expansion, the HT metric is expected to lose accuracy
for rapid spin rates, but the limits of validity of the approximation
are not well understood.  Weber \& Glendenning (1992) compared the
slow-rotation expansion with early numerical results by Friedman {\it
et al.} (1986). Their comparison led to `compatible results' down to
rotational Kepler periods $P_K\simeq 0.5$ ms. However their
self-consistency equation [Eq.\ (7) in Weber \& Glendenning (1992)] to
determine the mass-shedding frequency is not consistently truncated at
second order in angular velocity, as it should be. This is presumably
one of the reasons their method becomes very inaccurate close to the
limiting-mass model.

Numerical solutions for uniformly rotating stars can now be obtained
in both the slow and fast-rotation regimes. The HT approximation is
quite simple to implement: our own numerical code will be described in
Section \ref{integ}. In contrast, a solution of the Einstein equations
for rapidly rotating models is non-trivial.  Komatsu, Eriguchi and
Hachisu (1989) devised a self-consistent method to solve the full
Einstein equations, which was later improved by Cook, Shapiro and
Teukolsky (1994, henceforth CST). Stergioulas' RNS code (Stergioulas
\& Friedman 1995) is a variant of the CST code which has been extended
to allow for strange-star matter equations of state (EOSs),
differential rotation, computation of multipole moments and so on. A
very accurate spectral method for computing rapidly rotating stellar
models was independently developed by Bonazzola {\it et al.}
(cf. Bonazzola {\it et al.} 1993, Salgado {\it et al.} 1994) and
became part of the numerical relativity library LORENE (Bonazzola {\it
et al.} 1998). Nozawa {\it et al.}  (1998) showed that the different
numerical methods are in remarkable agreement, typical differences
being of order $10^{-3}$ or smaller for smooth EOSs. A new
multi-domain spectral method has been introduced by Ansorg {\it et
al.} (2002), who achieved near-machine accuracy for uniform density
stars; their results are in very good agreement with those of the
other groups.  The RNS code therefore provides a numerically accurate
description of fast-rotating relativistic stars.

Berti \& Stergioulas (2003, henceforth BS) used the RNS code to
compute constant rest-mass (so-called evolutionary) sequences of three
types using five realistic EOSs. The first type mimics the
evolutionary track of a canonical neutron star having gravitational
mass $M=1.4 M_\odot$ in the non-rotating limit. The second type has
the maximum gravitational mass $M=M_{\rm max}$ allowed by the given
EOS in the non-rotating limit. In the following, we will sometimes
refer to these sequences as type `14' and type `MM'.  The third type
is supramassive: its baryon mass is larger than allowed by the
non-rotating Tolman-Oppenheimer-Volkoff (TOV) equations. A
supramassive sequence does not terminate with a non-rotating model, so
that a comparison with HT models is not always possible. Therefore we
do not consider supramassive models in the following.

Another method of approximating the exterior spacetime takes the form
of a multipole expansion far from the star. This technique has been
used by Shibata \& Sasaki (1998) to derive approximate analytic
expressions for the location of the Innermost Stable Circular Orbits
(ISCOs) around rapidly rotating relativistic stars. However the
technique cannot easily be applied to other problems. Furthermore,
the relevant multipole moments must be computed {\it numerically} for
each model one wants to analyze, and then plugged into the relevant
formulas.

Approximate analytic models for fast rotating stars which include only
the most relevant multipolar structure of the exterior spacetime (that
is, the lowest-order mass and current multipoles) would be very useful
for astrophysical applications. In this paper we focus on an
interesting family of solutions found by Manko {\it et al.} (2000a,
2000b).  They used Sibgatullin's method (Sibgatullin 1991), which has
the advantage that the multipolar structure of the spacetime can be
prescribed \textit{a priori}, to find an exact, nine-parameter,
analytic solution of the Einstein-Maxwell equations. Consideration of
the full Einstein-Maxwell system is motivated by the fact that compact
stars normally have extremely large magnetic fields, which could
significantly change their properties (Bocquet {\it et al.} 1995) and
the development of instabilities (Spruit {\it et al.}  1999; Rezzolla
{\it et al.} 2000). However, for simplicity, we do not consider
effects due to magnetic fields in this paper.

An approximate description of a rapidly rotating neutron star can be
obtained using a five parameter sub-case of the analytic solution
(Manko 2000b). Ignoring charge and magnetic moment the free parameters
reduce to three. This three-parameter solution is an {\it exact}
solution of the Einstein equations in vacuum, but it only provides an
approximate description of a neutron star exterior\footnote{BS focused
on four out of the {\it a priori} infinite multipole moments of the
Manko metric: gravitational mass $M$, angular momentum $J$, mass
quadrupole moment $Q$ and current octupole moment $S_3$.  However
their statement that these are the only nonvanishing moments is not
precise: there do exist higher-order nonvanishing physical moments of
the Manko metric.}.

BS showed that the Manko solution approximates rather well the field
of a rapidly rotating neutron star in the fast rotation regime. They
used the three parameters of the analytic solution to match the
dominant multipole moments ($M$, $J$ and $M_2=-Q$) to given numerical
models. They found that an indicative measure of the deviation of the
Manko metric from the numerical metric is given by the relative error
in the first non-freely adjustable multipole moment: the current
octupole $S_3$.  BS found a peculiar feature of the Manko solution:
the solution's quadrupole moment {\it cannot} be matched to the
quadrupole moment of numerically computed spacetimes for slow rotation
rates. Therefore, it is of interest to understand whether the regimes
where the HT and the Manko metrics are applicable are completely
distinct, or have some range of $\Omega$ in common.

This paper compares the HT slow rotation expansion and the exact
vacuum solution by Manko {\it et al.} with the full
general-relativistic numerical CST spacetime. We will assume that our
numerical models, obtained using the RNS code, are {\it exact} (within
their numerical accuracy, that is of the order of $10^{-4}$ in each
computed quantity). An overview of the three different spacetimes we
consider in this paper is presented in Section \ref{intrometrics},
where we also introduce our notation.

The rest of the paper consists of two parts. In the first we integrate
the HT equations of structure for the same set of `realistic', nuclear
physics motivated EOSs examined in BS (Section \ref{integ}). From the
stable models in the HT sequence we choose those which have the same
gravitational mass and angular momentum as a given numerical
model. The procedure to do this is developed in Section \ref{match}.

In Section \ref{accurate} we quantify deviations of the HT and CST
spacetimes. A trustworthy measure of the deviation is given by the
relative error in the first non-freely adjustable multipole moment in
the HT spacetime: $M_2=-Q$. Picking a representative EOS we show that
deviations in $Q$ are at most $\sim 20 \%$ at the rotation rates
corresponding to the fastest observed millisecond pulsar PSR
J1939+2134 (Backer {\it et al.} 1982), for both mass sequences. We
also use some recent results by Abramowicz {\it et al.}  (2003) to
compute the ISCO radii predicted by the HT approximation for sequences
of slowly-rotating models. We find an interesting result: the HT
approximation gives {\it an excellent approximation to the ISCO
predicted by numerical solutions}, relative errors being smaller than
1~\% for astrophysically relevant rotation rates, down to the $1.5$~ms
period of PSR J1939+2134.

In the second part of the paper we compare the three vacuum exterior
spacetimes, focusing on invariant quantities, and looking at their
Petrov classification. In a nutshell, the Petrov classification of a
spacetime depends on which of the Weyl scalars can be set to zero by
suitable tetrad rotations (Chandrasekhar 1983, Stephani {\it et al.}
2003, Stephani 2004). The Kerr metric is of Petrov Type D, hence
algebraically special. In general, rotating perfect fluid spacetimes
will {\it not} be of Type D (see eg. Fodor and Perj\`es 2000), but
will be of (general) Type I. Our invariant comparison of the three
different models involves quantifying the deviation of each metric
from speciality. Indeed one might wish to assume (for example in
perturbative calculations of gravitational wave emission) that
rotating neutron star spacetimes are in some sense `close' to Type
D. Then one should develop a tool to quantify their `non-Type-D-ness':
here we use the curvature invariant speciality index $S$ introduced by
Baker \& Campanelli (2000). Each metric is expressed in different
coordinates, so we evaluate $S$ at the stellar radius, a physically
identifiable radial location. This quantity may also indicate whether
it is sensible to carry out perturbative calculations on neutron star
backgrounds in a `Type-D-approximation', and to extract the
gravitational wave content of the spacetime through a Teukolsky-like
formalism. In this paper, however, we will focus only on calculations
of the Weyl scalars in the background spacetimes.

We choose the simplest and most physically transparent tetrad to
compute the Weyl scalars and to characterise the non-Type-D-ness of
the spacetime. First of all we construct a transverse tetrad, in which
$\psi_1=\psi_3=0$, so getting rid of this tetrad gauge freedom
(Szekeres 1965). Then we carry out a tetrad rotation to go to the
canonical frame in which $\psi_0=\psi_4\equiv \psi_{04}$ (Pollney {\it
et al.} 2000; Re {\it et al.} 2003), that we call symmetric. In a Type
I spacetime, there are three distinct equivalence classes of
transverse frames (Beetle \& Burko 2002), each containing a symmetric
tetrad.  
From these three different symmetric tetrads we select a
`quasi-Kinnersley' tetrad (cf. Nerozzi {\it et al.} 2004), i.e. the
one that directly reduces to the canonical frame for Type D when we
set the parameters to reduce the metric to this type. 

We apply this procedure, which is described in detail in Section
\ref{tetrad}, to all three metrics.  In Section \ref{weyl} we show the
scalars and the corresponding speciality index. We find that
deviations from speciality on the equatorial plane are smaller than $5
\%$ at the stellar radius, even for the faster rotating models. These
deviations rapidly decrease for slower rotation rates and at larger
distances from the star. We find that, at leading order, the deviation
from Type D is proportional to $(Q-Q_{\rm Kerr})$. The conclusions
and a discussion follow.

\section{The three metrics}\label{intrometrics}

The three metrics we will consider in this paper (HT, Manko and CST)
can be written in the following general axisymmetric stationary form,
with coordinates $[t,c_1,c_2,\phi]$
\begin{eqnarray}
ds^2 &=&
g_{00}dt^2+2g_{03}dtd\phi+g_{33}d\phi^2+g_{11}dc_1^2+g_{22}dc_2^2
\label{genmetric}
\end{eqnarray}
where all the metric coefficients are functions of $c_1$ and $c_2$
only. Being stationary and axisymmetric the spacetime has two Killing
vectors, one timelike $\xi^a = [1,0,0,0]$ and one spacelike $\varphi^a
= [0,0,0,1]$.

The metrics can be explicitly written as:
\begin{eqnarray}
(ds^2)_{HT} &=& -e^{\nu}(1+2h)d t^2+e^{\lambda}[1+2m/(r-2M)]dr^2 \nonumber \\
& & {}+r^2(1+2k)[d\theta^2+\sin^2 \theta(d\phi-\omega dt)^2]
\end{eqnarray}
\begin{eqnarray}
(ds^2)_{Manko} &=& f^{-1}[e^{2\gamma}(d\rho^2+dz^2)+\rho^2d\tilde{\phi}^2] \nonumber \\
 & & {} -f(d\tilde{t}-\tilde{\omega} d\tilde{\phi})^2
\end{eqnarray}
\begin{eqnarray}
(ds^2)_{CST} &=& -e^{2\bar{\nu}}d\bar{t}^2+B^2\sin^2\bar{\theta}(d\bar{\phi}-\bar{\omega} d\bar{t})^2 \nonumber \\
 & & {} +e^{2\alpha}(d\bar{r}^2+\bar{r}^2 d\bar{\theta}^2)
\end{eqnarray}

Ignoring terms of ${\mathcal O}(\epsilon^3)$, the HT metric components
are (cf. Abramowicz {\it et al.} 2003)\footnote{The original Hartle
(1967) metric differs from the one presented here, in that it was not
consistently truncated to order $\epsilon^2$. The Abramowicz {\it et
al.\ } (2003) preprint contains some minor sign errors.}
\begin{eqnarray*}
g_{rr} & = & \hspace{2ex} \left(1-\frac{2M}{r}\right)^{-1}\left[1+j^2G_{1}+qF_{2}\right]+{\mathcal O}(\epsilon^3) \\
g_{tt} & = & - \left(1-\frac{2M}{r}\right)\left[1+j^2F_1-qF_{2}\right] +{\mathcal O}(\epsilon^3) \\
g_{\theta\theta} & = & \hspace{2ex} r^2\left[1+j^2 H_{1} - qH_{2}\right] +{\mathcal O}(\epsilon^3) \\
g_{\phi\phi} & = & g_{\theta\theta} \sin^2\theta +{\mathcal O}(\epsilon^3) \\
g_{t\phi} & = & \left(\frac{2jM^2}{r}\right)\sin^2\theta +{\mathcal O}(\epsilon^3) \\
\end{eqnarray*}
where
\begin{eqnarray}
F_1 &=& -pW+A_1, \nn\\
F_2 &=& 5r^3p(3u^2-1)(r-M)(2M^2+6Mr-3r^2) - A_1, \nn\\
A_1 &=& \frac{15r(r-2M)(1-3u^2)}{16M^2}\ln\left(\frac{r}{r-2M}\right), \nn\\
A_2 &=& \frac{15(r^2-2M^2)(3u^2-1)}{16M^2}\ln\left(\frac{r}{r-2M}\right), \nn\\
G_1 &=& p\left[(L-72M^5r)-3u^2(L-56M^5r) \right]-A_1, \nn\\
H_1 &=& A_2+(8Mr^4)^{-1}(1-3u^2)\times \nn\\
&\times&(16M^5+8M^4r-10M^2r^3+15Mr^4+15r^5), \nn\\
H_2 &=& -A_2+(8Mr)^{-1}5(1-3u^2)(2M^2-3Mr-3r^2), \nn\\
L &=& (80M^6\! +8M^4r^2\! +10M^3r^3+20M^2r^4-45Mr^5+15r^6), \nn\\
p &=& \left[8Mr^4(r-2M))\right]^{-1}, \nn\\
W &=& (r-M)(16M^5+8M^4r-10M^2r^3-30Mr^4+15r^5) \nn\\
&+&u^2(48M^6 - 8M^5r-24M^4r^2-30M^3r^3 \nn\\
&-&60M^2r^4+135Mr^5-45r^6), \nn
\end{eqnarray}
with $u=\cos\theta$. The dimensionless angular momentum and quadrupole
moment $j$ and $q$ are defined as:
\begin{equation}\label{jq}
j=J/M^2, \qquad q=Q/M^3.\nn
\end{equation}

The full Manko metric is rather lengthy and can be found in many
papers (eg. Manko \textit{et al}. 2000a, 2000b; Stute \& Camenzind
2002; BS), so we will not give it explicitly here. We only recall that
it depends on three parameters: the gravitational mass $M$, the
angular momentum per unit mass $a=J/M$ and a third parameter $b$ which
is related to the quadrupole moment by
\begin{equation}
Q=-M(d-\delta-ab+a^2),
\end{equation}
where 
\begin{eqnarray}
\delta&=&\frac{-M^2b^2}{M^2-(a-b)^2},\nn\\
d&=&\frac{1}{4 \left[M^2-(a-b)^2 \right]}.\nn
\end{eqnarray}
The next non-zero multipole moment for the Manko metric is the current
octupole, which is given by
\begin{equation}
S_3=
-M \left[a^3 -2a^2b+a [b^2+2(d-\delta)]-b(d-\delta) \right].
\end{equation}

For a detailed description of the CST metric we refer to the original
paper (Cook {\it et al.} 1994). The extraction of multipole moments
from the CST metric and the numerical calculation of ISCOs along
evolutionary sequences are discussed in BS.

In vacuum, the HT metric is equivalent to the Kerr metric at second
order in $J$ when one sets $Q=J^2/M$, and reduces to the Schwarzschild
solution when $J=Q=0$. It is not possible to obtain Kerr directly from
the Manko metric by a smooth variation of the real parameters $a$, $M$
and $b$. In other words, Kerr and Schwarzschild are not part of the 3
{\it real} parameter family of solutions described by the Manko
metric. However they can be obtained by analytic continuation, setting
$b^2=a^2-M^2$. Schwarzschild is a special case with $a=0$ and
$b^2=-M^2$.

\section{Integration of the Hartle-Thorne equations of structure}
\label{integ}

In this Section we outline some important aspects of the HT equations
of structure. The relevant formalism was originally developed in
(Hartle 1967), and the integration method is summarized in (Hartle \&
Thorne 1968). In the numerical code we used the reformulation of the
equations of structure given in Appendix A of (Sumiyoshi {\it et al.}
1999). We refer to that paper for the equations \footnote{Comparing
with the HT original papers, we found that Appendix A of (Sumiyoshi
{\it et al.} 1999) contains a few typos:
\begin{itemize}
\item In Equation (A11), $r^4$ should be replaced with $r^2 \xi$.
\item In Equation (A14), $R^4$ should be replaced with $r^4$.
\item In the last factor of Equation (A15), $r(r-2m)$ should be
replaced with $(r-2m)$.
\item In Equation (A36), $p$ should be replaced with $\rho_c$.
\item In Equation (A41), $e^{-\nu/2}$ should be replaced with $e^{\nu/2}$.
\item In Equation (A54), $\xi_0$ should be replaced with $\xi_2$.
\end{itemize}
Notice also that Sumiyoshi {\it et al.} do not explicitly give
boundary conditions at the center for the `particular' solutions
$h_2^p$, $v_2^p$. These boundary conditions are given in equations
(128)-(130) of Hartle (1967); however, in equation (130) of Hartle's
paper the factor $4\pi$ should be replaced by $2\pi$.}.

As discussed in the Introduction, the HT metric is a slow rotation
expansion in the angular velocity truncated to second-order.  We
consider the expansion parameter to be
\begin{equation}
\epsilon=\Omega/\Omega^{*}\;.
\end{equation}
The solutions of the TOV equations for a given EOS form a
one-parameter family, where, for example, one can use the star's
central energy density $\rho_c$ as the parameter. Similarly, solutions
of the HT equations of structure form a two-parameter family. Natural
parameters to label each model are the central energy density $\rho_c$
and the rotational parameter $\epsilon$.

Arriving at a particular HT model involves three conceptual
stages. First, the stellar structure of a non-rotating star is
obtained from the TOV equations, giving the non-rotating gravitational
mass $M$ and stellar radius $R$. Note that these non-rotating
quantities are unlabelled.  From these the rotational scale
$\Omega^{*}=\sqrt{M/R^3}$ can be calculated.  We can also compute the
star's binding energy $E_B$, defined as the difference between the
baryonic mass $M_b$ and the gravitational mass $M$:
\begin{equation}
E_B=M_b-M.
\end{equation}
Second, a model with $\epsilon=1$, i.e. a model rotating at the
`maximum' angular velocity of $\Omega^{*}$, is obtained from the full
HT equations.  Here all quantities are labelled with a superscript
asterisk : the model has gravitational mass $M^{*}=M+\delta M^{*}$,
angular momentum $J^{*}$, and quadrupole moment $Q^{*}$.  Finally, the
model with the chosen value of $\epsilon<1$ is calculated, having
rotating gravitational mass $M^{(\epsilon)}=M+\delta M^{(\epsilon)}$,
angular momentum $J^{(\epsilon)}$, and quadrupole moment
$Q^{(\epsilon)}$.

The effects of rotation on the various quantities of interest appear
at different orders in $\epsilon$. The general relativistic frame
dragging appears at first order, and the angular momentum is:
\begin{equation}\label{Jeq}
J^{(\epsilon)}=\epsilon J^{*}.
\end{equation}

The radius of gyration $R_g$, defined so that the moment of inertia is
$I=MR_g^2$, can then be calculated as
\begin{equation}
R_g=\left(\frac{I}{M}\right)^{1/2}=\left(\frac{J}{\Omega M}\right)^{1/2}
=\left(\frac{J^*}{\Omega^* M} \right)^{1/2} +{\cal O}(\epsilon^2)\;,
\end{equation}
where the first two relations are exact, and the third expresses the
fact that computing $R_g$ requires the integration of the equations of
structure at first order in $\epsilon$. However, involving the ratio
of $J$ and $\Omega$ (which are both proportional to $\epsilon$) $R_g$
does not depend on $\epsilon$. For the same reason, computing the
second order contribution to $R_g$ would require third order
corrections to $J$ and $\Omega$.

At second order in $\epsilon$ there appear corrections to the stellar
eccentricity $e_s$, quadrupole moment $Q$, mass $M$, binding energy
$E_B$, and coordinate stellar radius $R$. The stellar eccentricity is
\begin{equation}
e_s=\left[(R_e/R_p)^2-1\right]^{1/2}\,,
\end{equation}
where $R_e$ and $R_p$ are the stellar coordinate radii at the equator
and at the pole, respectively\footnote{In the notation of Hartle \&
Thorne (1968) the stellar radius
$R(\theta)=R+\xi_0(R)+\xi_2(R)P_2(\cos \theta)$, where $R$ is the
non-rotating stellar radius, $\xi_0(r)$ and $\xi_2(r)$ are radial
functions determined by the HT perturbation equations and $P_2(\cos
\theta)$ is a Legendre polynomial. The equatorial radius
$R_e=R(\pi/2)$, and the polar radius $R_p=R(0)$.}.
The quadrupole moment $Q$ can be invariantly defined as the leading
coefficient in the asymptotic behavior at large distances from the
origin of the eccentricity of a family of spheroidal surfaces where
the norm of the timelike Killing vector is constant.
Taking into account that the zero order (non-rotating) values of $e_s$
and $Q$ are zero, we have
\begin{eqnarray}
e_s^{(\epsilon)}&=&\epsilon^2 e_s^{*},\nn\\
Q^{(\epsilon)}&=&\epsilon^2 Q^{*},\nn\\
\delta M^{(\epsilon)}&=&\epsilon^2 \delta M^{*},\\
\delta E_B^{(\epsilon)}&=&\epsilon^2 \delta E_B^{*},\nn\\
\delta R^{(\epsilon)}&=&\epsilon^2 \delta R^{*}.\nn
\end{eqnarray}

This means, in particular, that for a fixed value of the central
energy density, a stellar model rotating with angular velocity
$\Omega=\epsilon \Omega^{*}$ has a gravitational mass
\begin{eqnarray}\label{gravm}
M^{(\epsilon)}=M+\epsilon^2 \delta M^{*}.
\end{eqnarray}

We have checked our numerical code by reproducing the results shown in
Hartle \& Friedman (1975) for polytropic EOSs.  We found agreement on
all significant digits for the quantities we have just defined. Then
we chose the same set of `realistic' EOSs used in BS, and verified
that in the non-rotating limit our numerical results are consistent
with Stergioulas' RNS code (described in Stergioulas \& Friedman
1995).

For any given `realistic' EOS, we can construct a sequence of rotating
models by integrating the HT structure equations for a series of
values of the central energy density. Our results are summarized in
Tables \ref{A}-\ref{APRb}.  The chosen central energy density
$\rho_c$, radius $R$, gravitational mass $M$, binding energy $E_B$ and
$\Omega^{*}$ are listed in the first five columns (marked by
$\epsilon^0$, since they do not depend on rotational corrections). At
order $\epsilon$ we list the value of the radius of gyration $R_g^{*}$
(from which the angular momentum $J^{*}$ can be easily obtained) and
of the frame dragging function $\omega_c^{*}$ at the center
(normalized by $\Omega^{*}$). Finally we list the second-order
quantities $\delta R^{*}/R$, $\delta M^{*}/M$, $\delta E_B^{*}/E_B$,
$Q^{*}/(MR^2)$ and $e_s^{*}$.

Let us stress again that although $\Omega^{*}$ is not the
mass-shedding limit in a general relativistic framework, because of
its physical meaning it does give an order of magnitude estimate of
the mass shedding frequency, and so we can expect the HT treatment to
break down when $\epsilon\sim 1$.

\subsection{Matching the Hartle-Thorne and numerical spacetimes}
\label{match}

To find a HT model that corresponds to a particular CST stellar model,
say one belonging to the constant-rest mass sequences computed in BS
using the RNS code, we adopt the following procedure, in which
quantities obtained from the CST model are labeled with overbars :
$\bar M$, $\bar J$ and $\bar Q$.

The aim is to find a HT model for which $M^{(\epsilon)}=\bar M$ and
$J^{(\epsilon)}=\bar J$.  Imposing these values on Eqs.  (\ref{Jeq})
and (\ref{gravm}) gives
\begin{eqnarray}\label{JJbar}
\epsilon=\bar J/J^{*}
\end{eqnarray}
and
\begin{eqnarray}\label{matchHT}
 \bar M =M+(\bar J/J^{*})^2 \delta M^{*}.
\end{eqnarray}
In (\ref{matchHT}), the quantities $J^{*}$, $M$ and $\delta M^{*}$ are
effectively functions of $\rho_c$, since once $\rho_c$ is specified
they can be calculated from $\epsilon=0$ or $\epsilon=1$
models. Therefore, in order to find the matching HT model, we must
search for a value of $\rho_c$ for which Eq. (\ref{matchHT}) holds;
$\epsilon$ is then given by Eq. (\ref{JJbar}).

In practice, we consider the function
\begin{eqnarray}\label{fzero}
f(\rho_c)=\bar M-\left[M+(\bar J/J^{*})^2 \delta M^{*}\right]\,.
\end{eqnarray}
To locate the matching model with a given accuracy (that we set to
$10^{-5}$) we search for the root of Eq. (\ref{fzero}) using
bisection. Note that, since the HT expansion is a small-$\epsilon$
approximation, we expect to find solutions only for small values of
the angular momentum $\bar J$. This turns out to be indeed the case.

In Table \ref{Matching} we show some basic features of the matching
models for the five EOSs we consider. The HT models we compute are
matched to the slowest-rotating models of the evolutionary sequences
presented in Tables 1-5 of BS: as just mentioned, Eq. (\ref{fzero})
has no solutions beyond some critical rotation rate.

For each EOS we present matching models for the 14 sequence above
matching models for the MM sequence. From left to right we list: the
central energy density $\bar \rho_c$ for which we find a matching
solution (if such a central energy density exists); the value of
$\epsilon=\bar J/J^{*}$ at the given central energy density; the
gravitational mass $\bar M$ and the angular momentum $\bar J$, in
geometrical units; the value of the quadrupole moment predicted by the
HT equations of structure for the selected model, which scales as
$\epsilon^2$, and is given by
\begin{eqnarray}
-M_2=Q^{(\epsilon)}=(\bar J/J^{*})^2 Q^{*};
\end{eqnarray}
and finally,
\begin{eqnarray}
\delta Q = 100(Q^{(\epsilon)}-\bar Q)/\bar Q,
\end{eqnarray}
that is, the percentage deviation of the HT quadrupole moment from its
value $\bar Q$ for the numerical spacetime, which can be found in
Tables 1-5 of BS.

\section{How accurate is the slow-rotation expansion?}
\label{accurate}

In the previous Section we have seen how, given a numerical solution
of the full Einstein equations with gravitational mass $M=\bar M$ and
(small enough) angular momentum $J=\bar J$, we can construct a HT
model matching the given values of $M$ and $J$. Our purpose in this
Section is to give two alternative, quantitative estimates of the
difference between each of the HT matching models and the
corresponding numerical model. The first estimate will be based on the
calculation of the spacetimes' quadrupole moment as a function of the
rotation rate; the second, on the calculation of the ISCO radii.

\subsection{Deviations in the quadrupole moment}
\label{quad}

The matching models described in Section \ref{match} are constructed
by imposing that the first mass multipole $M$ and the first current
multipole $J$ be the same as in the numerical spacetime. The HT
solution is a two-parameter family: once $\rho_c$ and $\epsilon$ are
specified, the whole multipolar structure of the HT spacetime is. In
particular, at second order in $\epsilon$ the HT structure equations
predict a specific value for the quadrupole moment $Q^{\rm HT}$. Since
multipole moments are global features of a spacetime (Fodor,
Hoenselaers \& Perj\`es 1989), we can measure the `distance' of the HT
spacetime from the numerical spacetime by looking at the difference
between the value $Q^{\rm HT}$ predicted by the HT code, and the
`exact' value $Q$ predicted by the RNS code. That is, we compute the
relative deviation
\begin{equation}
\frac{\Delta Q}{Q}=\frac{Q^{HT}-Q}{Q}.
\end{equation}
Given that the HT metric is exact in the limit $\epsilon \to 0$, but
becomes less accurate for larger values of $\epsilon$, we expect this
deviation to increase with the rotation rate: an explicit calculation
shows that this is indeed the case.

As shown in Table \ref{Matching}, matching models -- that is,
solutions of Eq.\ (\ref{fzero}) -- only exist up to some critical
value of the angular momentum $J$ (and of the expansion parameter
$\epsilon$) where the HT approximation breaks down, and is unable to
reproduce the numerical results. The deviation of $Q^{\rm HT}$ from
the `true' numerical value when $\epsilon\simeq 0.2$ depends slightly
on the EOS, but it is typically $\sim 10 \%$ for the 14 sequence and
$\sim 20 \%$ for the MM sequence. For the 14 sequence, the relative
error becomes larger than 20 \% when $\epsilon\sim 0.4$. From Table
\ref{Matching} one can see that deviations are not very sensitive to
the EOS: for example, when $\epsilon\sim 0.4$ the relative error for
the MM sequence ranges (roughly) between 25 \% and 30 \%.
At fixed $\epsilon$, $\Delta Q/Q$ is larger for the MM sequence than
for the 14 sequence. However recall that $Q$ is generally larger for
low-mass evolutionary sequences, where gravity is weaker and
centrifugal effects give the star a more oblate shape. Deviations in
the ISCO radii, as given by Eq.\ (\ref{AAKTeq}) below, depend on the
absolute value of $Q$. Since $Q$ is larger for the 14 sequence, and
percentage errors induced by the HT approximation are comparable to
those for the MM sequence, the HT approximation should induce larger
errors in the ISCO for the low-mass 14 sequence. This will be
explicitly shown in Section \ref{ISCO}.

\begin{figure}
\centerline{\resizebox{9cm}{!}{\rotatebox{0}{\includegraphics{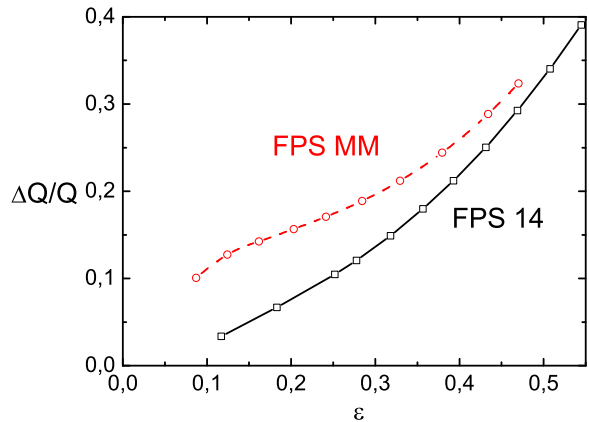}}}}
\caption{ 
Relative error in the quadrupole moment for the FPS 14 sequence
(squares) and the FPS MM sequence (circles).
}
\label{DeltaQ}
\end{figure}

Table 6 contains only a few slow-rotating models for each EOS. To get
a better idea of what happens in the slow-rotation regime we have
selected EOS FPS (which is quite average in terms of stiffness) and
computed more numerical models for small values of $J$ (that is, for
small values of $\epsilon$). Namely, we used the RNS code to compute
11 constant rest-mass models for the 14 sequence and 10 models for the
MM sequence, spanning the range $\epsilon=0$ to $\epsilon\sim 0.5$. To
each of these models we matched a HT model. Results for the relative
deviations in the quadrupole moment are shown in Fig.\
\ref{DeltaQ}. The convergence to zero as $\epsilon \to 0$ is very
smooth for the FPS 14 sequence. For the FPS MM sequence, the slowest
rotating model seems to show a slight deviation from the general
trend. We suspect this is due to a loss of accuracy of the RNS code in
computing the comparatively smaller $Q$ (for any given $\epsilon$) of
the more compact MM star models with small rotation rates. Indeed,
when we compute the curvature invariants for these models we also find
some numerical noise: this can be seen by a close scrutiny of the
bottom curve in the lower panels of Figs. \ref{fig6}, \ref{fig7} and
\ref{fig5}. The relative error $\Delta Q/Q$ is of the same order of
magnitude in both cases, but it is systematically larger for the FPS
MM sequence than for the FPS 14 sequence.

\subsection{Deviations in Innermost Stable Circular Orbits }
\label{ISCO}

The exterior vacuum metric for a HT model truncated at second order in
$\epsilon$  depends only on $M$, $J$ and $Q=Q^{\rm HT}$. In particular,
once we have computed these multipole moments we can use them to
compute the coordinate ISCO radius using the results by Abramowicz
{\it et al.} (2003). Using the dimensionless quantities $j$ and $q$
introduced in Eq.\ (\ref{jq}), their prediction for the {\it
coordinate} ISCO radii is
\beq\label{AAKTeq}
r_{\pm}&=&6M\left[
1\mp j\left(\frac{2}{3}\right)^{3/2}
+j^2\left(\frac{251647}{2592}-240\ln\frac{3}{2}\right)
\right.
\nn\\
&+&\left. q\left(-\frac{9325}{96}+240\ln\frac{3}{2}\right)
\right],
\eeq
where $r_+$ refers to corotating orbits, and $r_-$ to counterrotating
orbits. The comparison of the HT ISCO radii as given by
Eq.\ (\ref{AAKTeq}) to numerical results from the RNS code provides an
alternative measure of the difference between a numerical model and
the corresponding HT matching model.

We consider again the slow-rotating evolutionary sequence for EOS FPS
described in Section \ref{quad}. For each model along this
evolutionary sequence we compute $M$, $J$ and $Q^{\rm HT}$, and we
plug those values in Eq.\ (\ref{AAKTeq}) to compute the {\it
coordinate} ISCO radii. Then we can easily compute the {\it
circumferential} ISCO radii for these equatorial orbits as
\be
R^{\rm (HT)}_\pm=\sqrt{g_{\phi\phi}(r_\pm,\theta=\pi/2)},
\ee
and compare our results with the numerical value of the
circumferential ISCO radii\footnote{These are given by
$R_{\pm}=R_e+h_{\pm}$ in the notation of CST and BS.} as obtained from
RNS.

\begin{figure} 
{
\centerline{\resizebox{9cm}{!}{\rotatebox{0}{\includegraphics{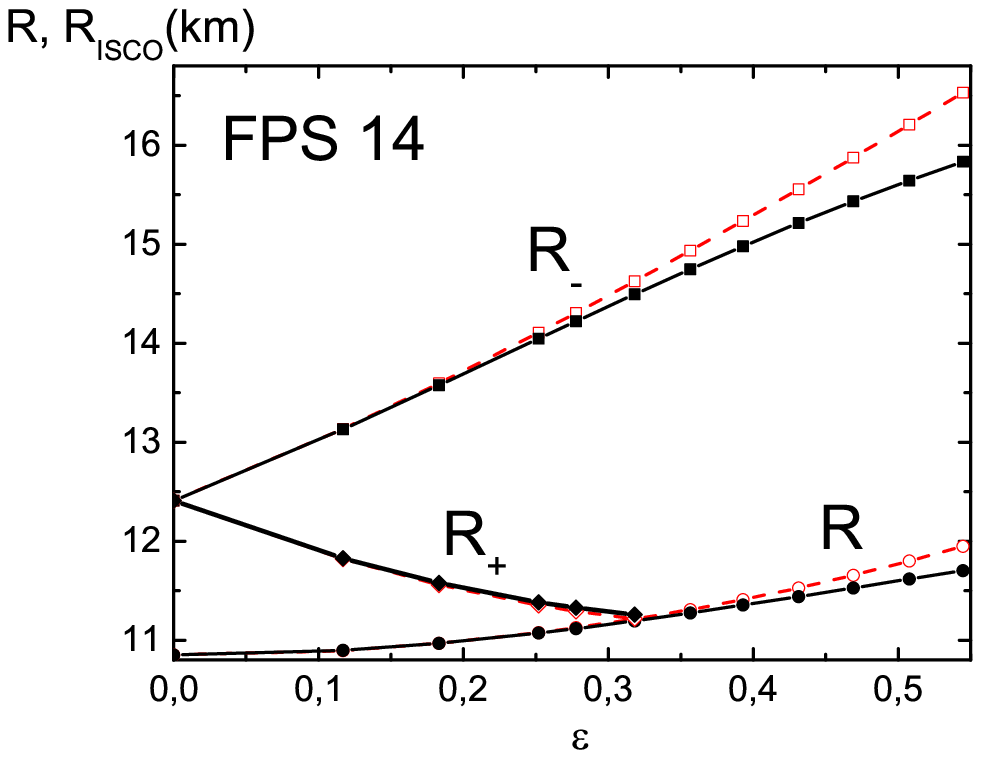}}}}
\centerline{\resizebox{9cm}{!}{\rotatebox{0}{\includegraphics{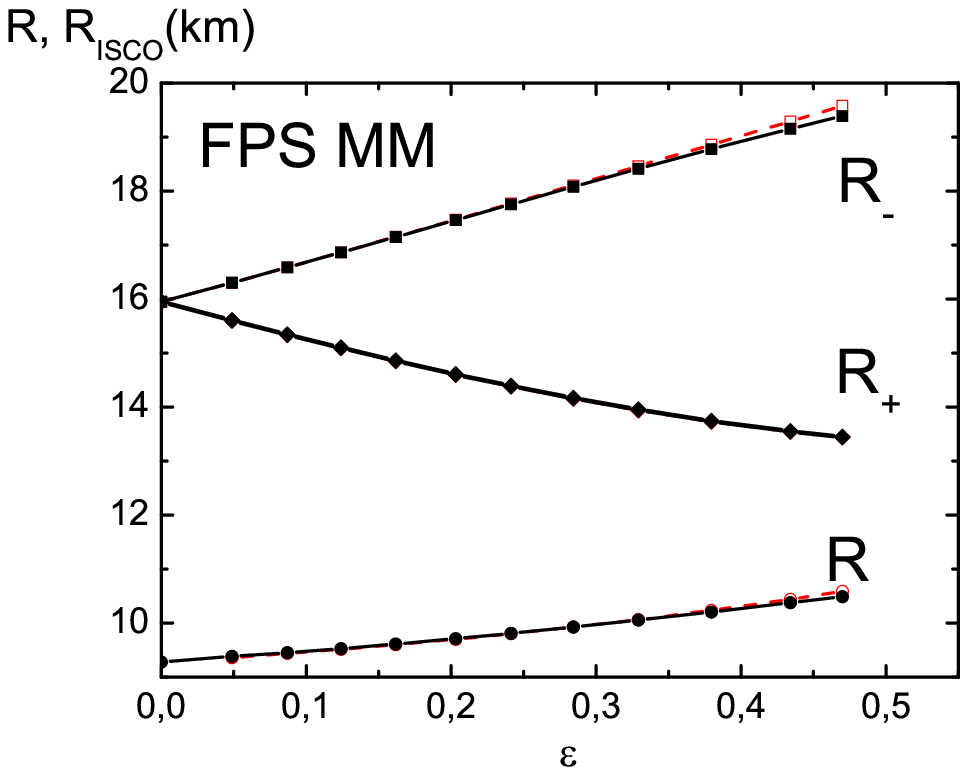}}}}
}
\caption{ 
Filled circles give the stellar circumferential radius $R$, filled
diamonds the circumferential radius of corotating ISCOs $R_+$ (when it
exists) and filled squares the circumferential radius of
counterrotating ISCOs $R_-$ for the numerical spacetime. Empty symbols
give the corresponding quantities for the HT metric. The top panel
refers to the FPS 14 evolutionary sequence, the bottom panel to the
FPS MM sequence.
}
\label{fig2}
\end{figure}

In Fig.\ \ref{fig2} we show results for the FPS 14 sequence (top
panel) and for the FPS MM sequence (bottom panel). In each panel we
plot together, as functions of $\epsilon$: 1) the circumferential
radii $R$ of the stars, 2) the corotating ISCOs $R_{+}$, 3) the
counterrotating ISCOs $R_{-}$. Filled symbols are obtained from RNS,
and empty symbols correspond to the HT matching models. The
circumferential radius for each matching model is just
\begin{equation} 
R^{\rm (HT)}_{e,{\rm circ}}=\sqrt{g_{\phi\phi}(R_{e},\theta=\pi/2)}, 
\end{equation} 
where $R_e$ is the {\it coordinate} equatorial radius, obtained
by integrating the HT equations of structure.

It is evident from Fig. \ref{fig2} that for neutron stars of canonical
mass $M\simeq 1.4 M_\odot$ the deviation of HT from the full numerical
solution is larger. However, even in that case the HT approximation
gives very accurate predictions both for the stellar equatorial radii
and for the ISCO radii up to $\epsilon\simeq 0.3$. In particular, the
fastest rotating model in the FPS 14 sequence for which a corotating
ISCO exists has $\epsilon=0.318$ and a spin frequency in physical
units $\nu_{\rm phys}=\Omega_{\rm phys}/2\pi=599$~s$^{-1}$,
corresponding to an orbital period $P=1.67$~ms. Therefore our
calculation shows that the HT approximation accurately predicts ISCO
radii down to periods which are comparable to that of the fastest
observed millisecond pulsar, PSR J1939+2134, spinning at $P=1.56$~ms
(Backer {\it et al.}  1982). 

\begin{figure} 
{
\centerline{\resizebox{9cm}{!}{\rotatebox{0}{\includegraphics{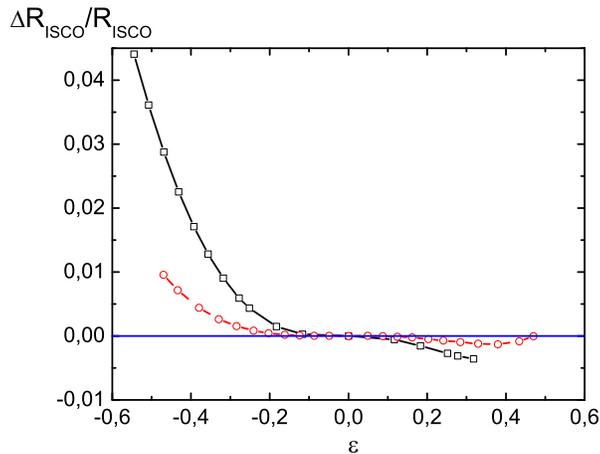}}}}
}
\caption{ 
Relative error in the ISCO radii, $(R^{(HT)}_+-R_+)/R_+$ and
$(R^{(HT)}_--R_-)/R_-$, for the FPS 14 sequence (squares) and the FPS
MM sequence (circles). In this plot we used the same convention as in
BS: a negative $\epsilon$ (and a negative $J$) corresponds to
counterrotating orbits.
}
\label{fig3}
\end{figure}

Relative errors for corotating and counterrotating ISCOs are shown in
Fig. \ref{fig3} for both the FPS 14 and FPS MM sequences. In this plot
we use the same convention as in BS: $\epsilon<0$ corresponds to
counterrotating orbits. As one would expect, errors tend to zero as
$\epsilon \to 0$. The error for counterrotating ISCOs monotonically
increases for large $\epsilon$; the error for corotating ISCOs does
not (at least for the FPS MM sequence) and it is typically extremely
small (below 5 parts in a thousand, roughly the same accuracy as our
numerical runs of the RNS code). We suspect that this could be due to
the second order HT approximation being sufficiently accurate to
compute the corotating ISCO, but not the counterrotating ones. To
confirm our conjecture we would need to push the HT calculation to
third order, which is beyond the scope of this paper.

As anticipated in Section \ref{quad}, ISCO radii for the MM sequence
agree better with the numerical results. Stars in the lower-mass FPS
14 evolutionary sequence have larger $Q$ (they are more oblate), and
relative errors in $Q$ for both sequences are roughly
comparable. Therefore effects due to the quadrupole moment on the
location of the ISCO are larger for the lower-mass sequence.  Notice
also that, although deviations in the quadrupole moment can be as
large as 20 \% for stars rotating with $\epsilon\sim 0.3$, the
relative errors on the ISCO are typically smaller than 1 \% at the
same rotation rates. This can be understood by looking at Eq.\
(\ref{AAKTeq}). For small rotation rates we have seen in Section
\ref{integ} that $q=a j^2$, where $a$ is a constant of order unity.
This scaling property is actually more general, applying also to
numerical solutions with large rotation rates: Laarakkers \& Poisson
(1999) showed that $a$ depends on the EOS and on the mass of the
stellar model, ranging between $a\simeq 2$ for high-mass models with
soft EOSs and $a\simeq 12$ for low-mass models with stiff EOSs
(cf. their Table 7; see also Miller 1977 for similar results in the
HT approximation). The numerical coefficient of the $j^2$-dependent
correction in Eq.\ (\ref{AAKTeq}) is $\simeq -0.226$, while the
numerical coefficient of the $q$-dependent correction is $\simeq
0.176$: that is, the two coefficients are roughly equal in magnitude
but opposite in sign. For this reason, shape-dependent effects in the
ISCO (which are ${\mathcal O}(\epsilon^2)$) tend to cancel for small
rotation rates. This explains why the HT approximation does such a
good job in predicting the ISCO radii.

\begin{figure} 
{
\centerline{\resizebox{9cm}{!}{\rotatebox{0}{\includegraphics{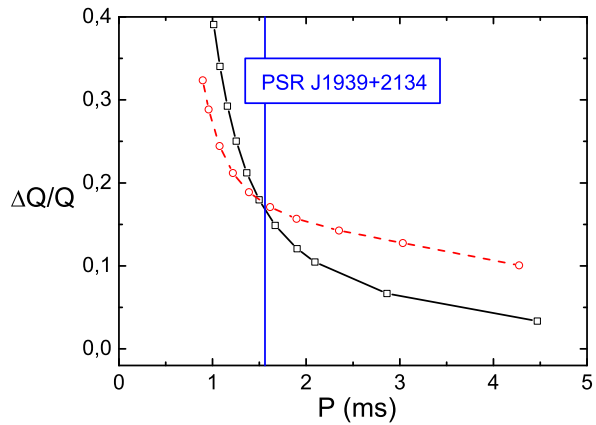}}}}
\centerline{\resizebox{9cm}{!}{\rotatebox{0}{\includegraphics{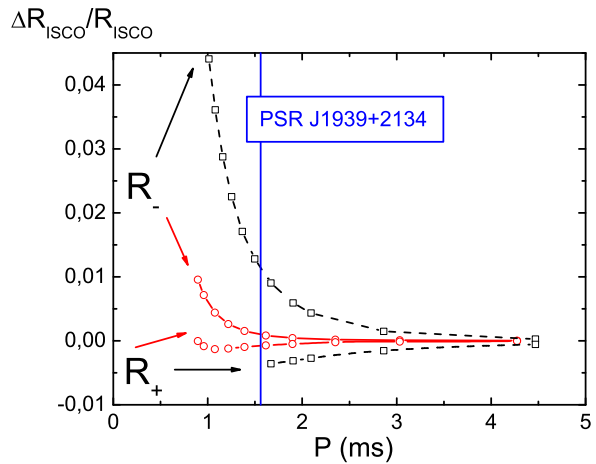}}}}
}
\caption{ 
The top panel shows the relative error in the quadrupole moment for the
FPS 14 sequence (squares) and the FPS MM sequence (circles) as a
function of the spin period $P$ of the star (in milliseconds).  The
bottom panel shows the relative error for corotating ($R_+$) and
counterrotating ($R_-$) ISCO radii for the FPS 14 sequence (squares)
and the FPS MM sequence (circles) as a function of $P$.  In both
panels, the vertical line marks the spin frequency of the fastest
observed millisecond pulsar PSR J1939+2134.
}
\label{Periods}
\end{figure}

To better illustrate the physical significance of our results, in
Fig. \ref{Periods} we plot the relative errors in the quadrupole
moments (top panel) and in the ISCO radii (bottom panel) as a function
of the spin period $P$ (in milliseconds) of our rotating stellar
models. This plot is useful for the following reasons. First of all,
the relation between the HT expansion parameter $\epsilon$ and the
rotational period of pulsars is non-linear. Not only is $P$
proportional to $1/\epsilon$, it also has (for the constant-rest mass
sequences we consider) a weak but non-trivial dependence on the
parameters $M$ and $R$:
\be
P=\frac{2\pi}{\Omega}=\frac{2\pi}{\epsilon \Omega^*}
=\frac{2\pi}{\epsilon (M/R^3)^{1/2}}\,.
\ee
Fig. \ref{Periods} shows that all of our so-called `slow rotation'
models are actually millisecond pulsars, many of which rotate even
faster than the fastest observed millisecond pulsar PSR J1939+2134
(Backer {\it et al.} 1982). This gives a concrete idea of the rotation
rates we are considering in this paper.

In both panels we mark by a vertical line the spin frequency of PSR
J1939+2134. The relative error in $Q$ for a pulsar spinning as fast as
PSR J1939+2134 is of order 20 \%, and this is quite independent of the
stellar mass. Remarkably, the relative error in corotating and
counterrotating ISCOs induced by the HT approximation is smaller than
1 \%, even when the pulsar rotates with a period as small as $1.5$
ms. This upper limit on deviations in the ISCO applies both to the FPS
14 sequence and to the FPS MM sequence.

For concreteness, in Fig. \ref{Periods} we presented the specific
example of PSR J1939+2134. However we wish to stress that our analysis
can be considered more general. Even if faster pulsars were
discovered, our conclusions are likely to be only marginally
altered. Indeed, Chakrabarti {\it et al.} (2003) performed a
statistical analysis of observations of 11 nuclear powered pulsars
with spin frequencies between 270 Hz and 619 Hz. Their analysis
implies an upper limit of 760 Hz ($95 \%$ confidence level) on the
spin frequency, to be compared with the 641 Hz of PSR
J1939+2134. Therefore PSR J1939+2134 should be very close to the {\it
maximum} spin frequency allowed for an isolated neutron
star. Interestingly, the upper limit from burst oscillations is much
lower than the mass-shedding limit predicted by most EOSs: some
mechanism must be at work to halt pulsar spin-up while accretion is
still active. A plausible explanation is gravitational wave emission,
due either to the excitation of the $r$-mode instability or to the
presence of an accretion-induced crustal quadrupole moment (see
Chakrabarti {\it et al.} 2003 for a discussion).

\section{Weyl scalars in a quasi-Kinnersley frame}
\label{tetrad}

In the remainder of this paper we concentrate on the vacuum
surrounding our rotating stellar models. We compute the Weyl scalars
for the HT, Manko and CST spacetimes and define the `quasi-Kinnersley'
frame (Section \ref{QK}). In Section \ref{HTQK} we present the simple
analytical results we obtained for the HT metric, and in Section
\ref{MCSTWS} we describe the analogous calculation for the Manko and
CST metrics. In Section \ref{weyl} we give numerical results for all
three spacetimes.

\subsection{The quasi-Kinnersley tetrad}
\label{QK}

We introduce a Newman-Penrose null tetrad \{$l^a$, $n^a$, $m^a$,
$\bar{m}^a$\}, where $l^a$ and $n^a$ are real and $m^a$ and
$\bar{m}^a$ are complex conjugate, satisfying the usual orthonormality
conditions: $l^an_a=1$, $m^a\bar{m}_a=-1$, all other products
zero. The Weyl scalars are the usual contractions of the Weyl tensor,
\begin{eqnarray}
\psi_0&=&-C_{pqrs} l^{p} m^{q} l^{r} m^{s}, \\
\psi_1&=& -C_{pqrs} l^{p} n^{q} l^{r} m^{s},\nn \\
\psi_2&=&-C_{pqrs} l^{p} m^{q} \bar{m}^{r} n^{s},\nn\\
\psi_3&=& -C_{pqrs} l^{p} n^{q} \bar{m}^{r} n^{s},\nn \\
\psi_4&=&-C_{pqrs} n^{p} \bar{m}^{q} n^{r} \bar{m}^{s}.\nn
\end{eqnarray}
The speciality index (Baker \& Campanelli 2000) is defined as
\begin{eqnarray}\label{SI}
 S=27J^2/I^3,
\end{eqnarray}
where the curvature invariants $I$ and $J$ can be expressed as
\begin{eqnarray*}
I&=&3(\psi_2)^2-4\psi_{1}\psi_{3}+\psi_{0}\psi_{4}, \\
J&=&-(\psi_2)^3+\psi_{0}\psi_{2}\psi_{4}+2\psi_{1}\psi_{2}\psi_{3}
  - \psi_{4}(\psi_{1})^2-\psi_{0}(\psi_{3})^2.
\end{eqnarray*}
We use the additional conventions that an NP tetrad with
$\psi_{1}=\psi_{3}=0$ is called transverse (Beetle \& Burko 2002), and
that if further $\psi_{0}=\psi_{4}$ then it is called symmetric, in
which case we may refer to $\psi_{0}=\psi_{4}$ as $\psi_{04}$.

The interpretation of the Weyl scalars in the context of BH
perturbation theory is that $\psi_{0}$ represents radiation along the
$l^a$ direction, $\psi_{4}$ represents radiation along the $n^a$
direction, while $\psi_{2}$ describes the Coulomb effect due to the
presence of a central mass and the frame dragging of the background
Kerr spacetime. $\psi_{1}$ and $\psi_{3}$ are longitudinal effects of
no physical interest that can be set to zero by the remaining tetrad
freedom (Teukolsky 1973, Stewart \& Walker 1974). In our case,
however, and in general for a stationary spacetime of Petrov Type I,
$\psi_{0}$ and $\psi_{4}$ represent transverse curvature deviations
(cf. Szekeres 1965) from the Coulombian $\psi_{2}$ field due to
rotation: see Eq. (\ref{psi04HT}).

It is useful then to consider only transverse frames. In a Type D
spacetime a frame can be found in which $\psi_{2}$ is the only
non-zero scalar: we refer to this as the Kinnersley frame
(cf. Teukolsky 1973).  With an appropriate choice of parameters (see
below) both the Manko and the HT metrics can be reduced to Type D
spacetimes (i.e.\ Schwarschild and Kerr). Here we work in the
symmetric tetrad which, under these parameter choices, becomes the
Kinnersley frame. In this tetrad $\psi_{04}$ is an effect due solely
to the non-Type-D-ness of the full metrics.  We call this tetrad the
`quasi-Kinnersley' tetrad (Nerozzi {\it et al.} 2004).

In a Type I spacetime, there are three distinct equivalence classes of
transverse tetrads (Beetle \& Burko 2002). Each class consists of all
those transverse tetrads that can be related to each other by either
the exchange freedom $l^a \leftrightarrow n^a$, or by a class III
rotation\footnote{The parameter $\theta$ used in this Section has
obviously nothing to do with the angular coordinate in the HT
metric.}: \\[0.3cm]
\begin{tabular}{lclcl}
$l \rightarrow  A^{-1}l$      & & $\psi_{0} \rightarrow A^{-2}e^{2i\theta}\psi_{0}$ & & $\psi_{3} \rightarrow Ae^{-i\theta}\psi_{3}$  \\
$n \rightarrow  A{}n$         & & $\psi_{1} \rightarrow A^{-1}e^{i\theta}\psi_{1}$  & & $\psi_{4} \rightarrow A^{2}e^{-2i\theta}\psi_{4}$ \\
$m \rightarrow  e^{i\theta}m$ & & $\psi_{2} \rightarrow \psi_{2}$ & &\\
\end{tabular}\\[0.2cm]

We find one particular transverse tetrad by making the ansatz that the
real null vectors of the tetrad should be linear combinations of the
Killing vectors $\xi^a$ and $\varphi^a$
\begin{eqnarray}
l^a &=& A\xi^a+B\varphi^a,\\
n^a &=& C\xi^a+E\varphi^a.
\end{eqnarray}
We then impose the NP tetrad inner product conditions and make
particular (arbitrary) choices of the remaining free parameters and
signs (different choices may result in a different tetrad, but this
would be related to the first by a class III rotation, and so would be
in the same equivalence class).  The result is
\begin{eqnarray}
l^a &=& \frac{1}{\sqrt{2}} \frac{\sqrt{g_{33}}}{\sigma} \left[1,0,0,\frac{(-g_{03}-\sigma)}{g_{33}} \right], \\
n^a &=& \frac{1}{\sqrt{2}} \frac{\sqrt{g_{33}}}{\sigma} \left[1,0,0,\frac{(-g_{03}+\sigma)}{g_{33}} \right], \\
m^a &=& \frac{1}{\sqrt{2}} \left[0,\frac{i}{\sqrt{g_{11}}},\frac{1}{\sqrt{g_{22}}} ,0 \right],
\end{eqnarray}
where
\begin{eqnarray}
\sigma^2 &=& g_{03}^2-g_{00}g_{33}.
\end{eqnarray}
To pick out a particular member of a class of tetrads one can choose
the  symmetric  tetrad.  This can always be found from a
transverse tetrad through the following class III rotation
\begin{eqnarray}
\label{radscal}
\hat{\psi}_2 = \psi_2, &\hspace{3ex}& \hat{\psi}_{04} \equiv \hat{\psi}_0 = \hat{\psi}_4 = \sqrt{\psi_0 \psi_4},\\
\Rightarrow \hspace{2ex} A^4 = |\psi_0| / |\psi_4| & , \hspace{3ex}&   
\theta = \frac{1}{4}\arctan\left[ 
\frac{\Im \left(\overline{\psi}_0 \psi_4 \right)}{\Re \left(\overline{\psi}_0 \psi_4 \right)}
\right],
\end{eqnarray}
where the $\hat{\psi}$'s are in the symmetric tetrad, the $\psi$'s are
in the non-symmetric tetrad and an overbar means complex conjugate. We
do not write down the expressions for the tetrad vectors in the
symmetric frame, since the parameters $A$ and $\theta$ are very long
if written out in full, making the expression unenlightening, but it
is worth noting that since this is a class III rotation, the
directions of the $l^a$ and $n^a$ vectors do not change.

We will call the above tetrad, built on the Killing vectors, T1. Using
a method suggested by Burko (2003) we can show that if the
non-zero scalars in the symmetric version of T1 are ${\psi^{\rm
T1}_{04}}$ and ${\psi^{\rm T1}_{2}}$ then in the other two transverse
\textit{symmetric} tetrads (T2 and T3) the non-zero scalars are given
by:
\renewcommand{\arraystretch}{2}
\begin{eqnarray}
\textbf{T2} \hspace{1cm} {\psi^{\rm T2}_{2}}  &=& \frac{1}{2} \left({\psi^{\rm T1}_{04}}-{\psi^{\rm T1}_{2}}\right) \\
                         {\psi^{\rm T2}_{04}} &=& \frac{1}{2} \left({\psi^{\rm T1}_{04}}+3{\psi^{\rm T1}_{2}}\right)\\
\textbf{T3} \hspace{1cm} {\psi^{\rm T3}_{2}}  &=& \frac{1}{2} \left({-\psi^{\rm T1}_{04}}-{\psi^{\rm T1}_{2}}\right) \\
                         {\psi^{\rm T3}_{04}} &=& \frac{1}{2} \left({\psi^{\rm T1}_{04}}-3{\psi^{\rm T1}_{2}}\right)
\end{eqnarray}

Hence, once the scalars in one transverse tetrad are known, it is easy
to calculate those in the other two frames. We have done this in the
reductions of Manko and HT to the Type D metrics described in the next
section. In each case it turns out that T3 is the Kinnersley tetrad,
i.e.  that frame in which, for the Type D metrics, $\psi_{2}$ is the
only non-zero scalar. We therefore make all of the following analysis
in T3. This means that we are making an analogous choice of tetrad in
each metric, that all tetrad freedom is used up, and that $\psi_{04}$
represents a pure Type I effect. Indeed, it is clear from the
definition (\ref{radscal}) that each of the $\psi_{04}$'s in each of
the three transverse symmetric tetrads is the square root of one of
the three corresponding curvature invariants (`radiation scalars')
introduced by Beetle \& Burko (2002)\footnote{Although each of the
three $\psi_{04}$'s is defined in a specific tetrad by
Eq. (\ref{radscal}), which is not {\it per-se} invariant under tetrad
rotations of Type I and II, it can actually be expressed in terms of
$I$ and $J$ only. See Beetle and Burko (2002) for more
details.}. However, only $\psi_{04}$ in the quasi-Kinnersley frame
vanishes in the `Type D limit', and therefore its value is an
invariant measure of the deviation from Type D.

Finally, in any symmetric transverse frame (in particular, in the
quasi-Kinnersley tetrad T3) Eq.\ (\ref{SI}) reduces to
\begin{eqnarray}\label{SIQK}
S &=& 27\left[(-\psi_{2}^3+\psi_{2}\psi_{04}^2)^2\right]
/\left[(3\psi_{2}^2+\psi_{04}^2)^3\right].
\end{eqnarray}
$S$ is equal to $1$ in a Type D spacetime. $(1-|S|)$ is also
dimensionless; in this sense it is better than $\psi_{04}$ as an
invariant measure of the deviation from Type D. We will refer to
$(1-|S|)$ as the Non-Type-D-ness in the following.

\subsection{The Hartle-Thorne Weyl scalars in the quasi-Kinnersley frame}
\label{HTQK}

The HT slow rotation parameter $\epsilon$ is {\it not} a perturbative
parameter with respect to Type D: at linear order in $\epsilon$ the HT
solution in vacuum is equivalent to the Kerr metric, which of course
is Type D {\it at all orders} in $\epsilon$. Discarding terms of
${\mathcal O}(\epsilon^3)$ and higher, the HT Weyl scalars in the
quasi-Kinnersley (symmetric transverse, T3) tetrad have simple
expressions:
\begin{eqnarray}
\label{psi04}
(\psi_{04})_{HT} &=&
\frac{15\sin^2\theta}{32M^7r^3(r-2M)}(Q-Q_{\rm Kerr})\times\nn\\
&\times&\left[3r^2(r-2M)^2\ln[r/(r-2M)]\right.\nn\\
&+&\left.2M(r-M)(2M^2+6rM-3r^2)\right]\,,
\label{psi04HT}
\end{eqnarray}
\begin{eqnarray}
(\psi_{2})_{HT} &=& -\frac{M}{r^3}-\frac{3i\cos\theta J}{r^4}\nn\\
&-&\frac{1}{16M^6r^7}\left\{P_2(\cos\theta)\left[15r^7\ln\left(\frac{r}{r-2M}\right) \right.\right.\nn\\
&-&10Mr^2(8M^4+6M^3r+4M^2r^2+3Mr^3+3r^4) \nn \\
&-&\left.\left.16M^6(r-6M)\right]-16M^5r(r-M)\right\}J^2\nn\\[0.2cm]
&+&\frac{1}{16M^5r^4}P_2(\cos\theta)\left[15r^4\ln\left(\frac{r}{r-2M}\right) \right. \nn \\
&-&\left.10M(3r^3+4rM^2+3r^2M+6M^3)\right]Q\,. 
\label{psi2HT}
\end{eqnarray}
For a Kerr BH $Q=Q_{\rm Kerr}=J^2/M$, and it is immediately seen that
$\psi_{04}=0$ (as it should). Writing the above relations as Taylor
series of the form
\begin{eqnarray}
\label{exp}
\psi_{2} &=& \psi_{2}^{(0)}+\psi_{2}^{(1)} \epsilon
+\psi_{2}^{(2)}\epsilon^2+{\mathcal O}(\epsilon^3)\,,\\
\psi_{04} &=& \psi_{04}^{(2)}\epsilon^2+{\mathcal O}(\epsilon^3)\,,\nn
\end{eqnarray}
it is easy to show that
\begin{equation}
\label{expS}
1-S=
3\left(\frac{\psi_{04}^{(2)}}{\psi_{2}^{(0)}}\right)^2 \epsilon^4
+{\mathcal O}(\epsilon^5)=
3\left(\frac{\psi_{04}}{\psi_{2}}\right)^2+{\mathcal O}(\epsilon^5)\,.
\end{equation}

Remarkably, the leading-order correction of $S$ around its Type D
value $S=1$ is only of order $\epsilon^4$: a fact due to $S$ being a
non-linear function of the Weyl scalars. Indeed, this doesn't mean
that perturbations of order $\epsilon^2$ do not alter the speciality
condition: rather, $(1-S)$ is quadratic in these
perturbations\footnote{It is easy to check, using an expansion like
(\ref{exp}) extended to higher order, that the leading fourth-order
term in (\ref{expS}) will not change.}. This is quite a subtle point,
that can better be understood by explicitly constructing the principal
null directions. Such a construction has been worked out in detail by
Cherubini {\it et al.}  (2004) for the simpler case of the vacuum
Kasner spacetimes: cf. in particular the discussion following their
Eq. (2.15).  Furthermore, at leading order in the expansion the
Non-Type-D indicators $1-S$ and $\psi_{04}/\psi_{2}$ are completely
equivalent to each other. In Section \ref{weyl} we show by explicit
calculations that this equivalence of the two Type D indicators is in
fact more general: it holds true (at least approximately) for fully
relativistic rotating star spacetimes as well.  From
Eq. (\ref{psi04HT}) we also see that $(\psi_{04})_{HT}$ is
proportional to $(Q-Q_{\rm Kerr})$.  In other words, at leading order
deviations from Type D are driven by the deviation of the quadrupole
moment from its Kerr value, as we would expect on physical grounds.

From Eqs. (\ref{psi04}) and (\ref{psi2HT}) the asymptotic behavior of
the Weyl scalars for large $r$ at leading order is:
\begin{eqnarray}
\label{psi04HTasy}
(\psi_{04})_{HT} &\sim& \frac{3 \sin^2 \theta (Q-Q_{\rm Kerr})}{2r^5}\,,\\
(\psi_{2})_{HT} &\sim& -\frac{M}{r^3}\,.
\end{eqnarray}
Consequently, for the variables we use to measure deviations from Type
D, at large $r$ we have:
\begin{eqnarray}
\left|\frac{\psi_{04}}{\psi_2}\right|
&\sim&\frac{3\sin^2(\theta)}{2Mr^2}(Q-Q_{\rm Kerr})\,,\\
1-S
&\sim&\frac{27\sin^4(\theta)}{4M^2r^4}(Q-Q_{\rm Kerr})^2\,.
\end{eqnarray}

Finally, as a useful check of our results, we show that $\psi_2$
reduces to the correct limit in the Kerr case. The proof is quite
straightforward.  To avoid confusion, let us denote the standard
Boyer-Lindquist coordinates for the Kerr metric as $(s,\alpha)$
instead of $(r,\theta)$. Then the only non-zero Weyl scalar for the
Kerr metric in the Kinnersley frame is given by
\begin{eqnarray}
(\psi_{2})_{\rm Kerr} &=& \frac{-M}{(s-ia\cos\alpha)^3}. \label{psi2Kerr}
\end{eqnarray}
The relation between the two sets of coordinates is\footnote{ For the
HT metric to reduce to the Kerr metric we must set $J=Ma$, $Q=J^2/M$
and ignore terms of ${\mathcal O}(J^3)$. Formula (A5) in (Hartle \&
Thorne 1968) contains a typo: in the $r$ transformation, $\cos^2
\theta$ should be replaced by $-\cos^2 \theta$. }:
\begin{eqnarray}
\label{HT(A5a)} s &=& r\left\{1-\frac{a^2}{2r^2}\left[\left(1+\frac{2M}{r}\right) \left(1-\frac{M}{r}\right)\right.\right.\\
&-&\left.\left.\cos^2\theta\left(1-\frac{2M}{r}\right)\left(1+\frac{3M}{r}\right)\right]\right\}, \nn \\
\alpha &=& \theta-a^2\cos\theta \sin\theta\frac{1}{2r^2}\left(1+\frac{2M}{r}\right). \label{HT(A5b)}
\end{eqnarray}
Carrying out these coordinate transformations on equation
(\ref{psi2Kerr}), setting $J=Ma$, $Q=J^2/M$, and discarding terms of
${\mathcal O}(J^3)$, we get
\begin{eqnarray}
(\psi_{2})_{\rm Kerr} &=& -\frac{M}{r^3}-\frac{3i\cos\theta J}{r^4}\\
&+&\frac{3}{2}\frac{(r-M)}{r^7M}\left[(6M+5r)\cos^2\theta-(2M+r)\right]J^2 \nn
\end{eqnarray}
This same expression is obtained from equation (\ref{psi2HT}) when
$Q=J^2/M$ is substituted.

This confirms that in T3, with the parameters set for Kerr, we indeed
have the Kinnersley frame, and that with these parameters, $\psi_2$ is
the only non-zero Weyl scalar and matches the analytically known
expression.  Clearly the further reduction to Schwarzschild ($J=Q=0$)
also gives the correct known expression for $(\psi_2)_{\rm Schw}$.  

\subsection{The Manko and CST Weyl scalars in the quasi-Kinnersley frame}
\label{MCSTWS}

For the Manko metric we computed the Weyl scalars in the
quasi-Kinnersley frame by the procedure described in Section \ref{QK},
using the algebraic manipulation program {\it Maple}. However the
final analytic expressions are very long and unenlightening, so we do
not give them here.  As a check, we verified numerically that in the
quasi-Kinnersley frame $(\psi_{04})_{Manko}$ is zero when the Manko
parameters are set to Kerr ($b^2=a^2-M^2$) or Schwarzschild ($a=0$,
$b^2=-M^2$).

For the CST metric we followed again the same procedure, but in this
case we had to solve a numerical problem. The Weyl scalars depend on
first and second order partial derivatives of the metric
coefficients. For the analytical spacetimes (HT and Manko) the
computation of the scalars is a trivial process. However, for the
numerical spacetime one has to invoke numerical differentiation.  The
second order radial derivatives turn out to be more sensitive to
numerical error. Therefore we find that it is advantageous to avoid
using second order radial derivatives calculated straight from a
second order finite difference scheme. Instead we use the Ricci
identities in the Newman-Penrose formalism. These allow us to express
all second order radial derivatives in terms of second order angular
derivatives and first order derivatives.  Once we do this, the
numerical calculation of the Weyl scalars turns out to be second order
convergent on our numerical grid.

\section{Non-Type-D-ness of the exterior vacuum}
\label{weyl}

Using the procedure described in the previous Section we can compute
the non-zero Weyl scalars $\psi_2$ and $\psi_{04}$ in the
quasi-Kinnersley frame for all three metrics (HT, Manko and CST). Then
we can evaluate the Non-Type-D-ness $(1-|S|)$ and the ratio
$|\psi_{04}/\psi_2|$. To avoid unnecessary complications, in the
following we will restrict consideration to the equatorial plane.

\begin{figure} 
{
\centerline{\resizebox{9cm}{!}{\rotatebox{0}{\includegraphics{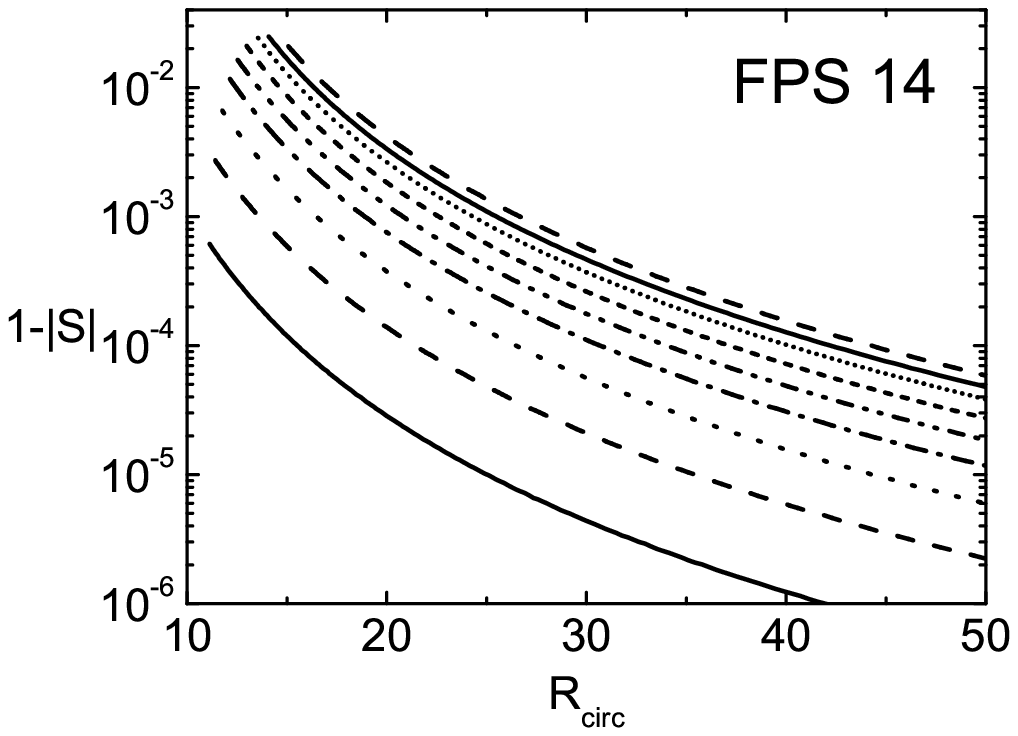}}}}
\centerline{\resizebox{9cm}{!}{\rotatebox{0}{\includegraphics{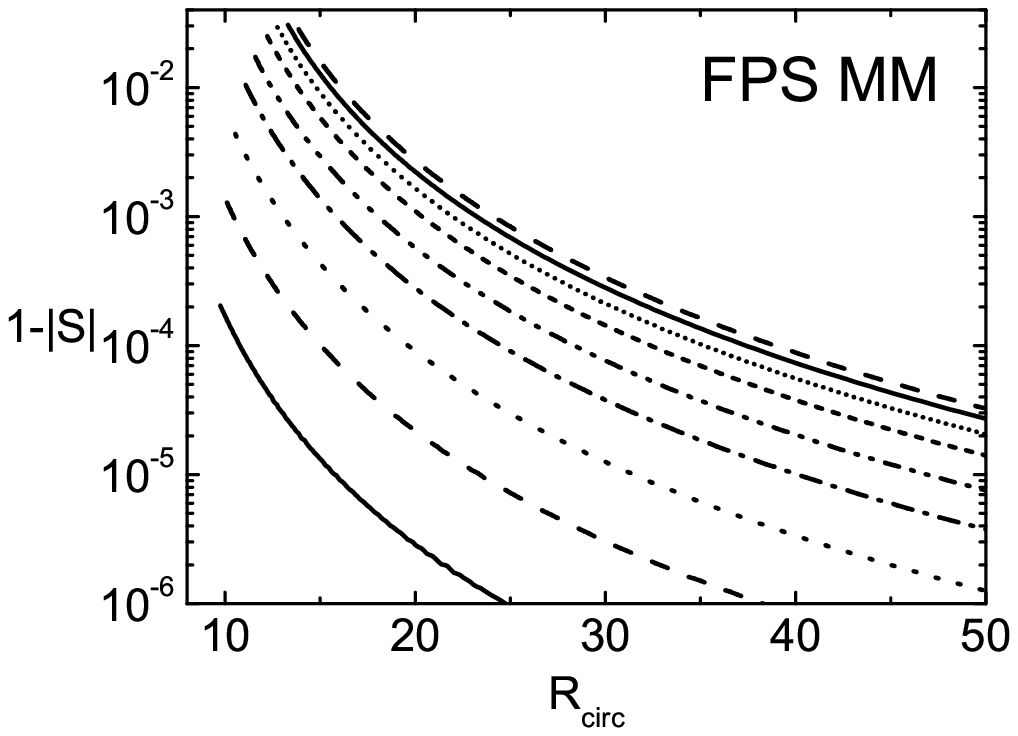}}}}
}
\caption{ Non-Type-D-ness $(1-|S|)$ for the FPS 14 (top panel) and FPS
MM (bottom panel) sequences, computed using the CST metric, as a
function of the circumferential radius $R_{\rm circ}$.  Within each
panel, lines from bottom to top give $(1-|S|)$ as a function of the
circumferential radius for different values of the rotation rate:
slower rotating models are closer to Type D, so $(1-|S|)$ is
smaller. Since the calculation only applies to the vacuum exterior
spacetime, each line starts from the star's circumferential equatorial
radius $R_e$.  }
\label{fig6}
\end{figure}

In Fig.\ \ref{fig6} we plot the invariant Non-Type-D-ness $(1-|S|)$ as
a function of the circumferential radius $R_{\rm circ}$ for the CST
metric. Throughout this section we will always refer to
circumferential radii, which are operationally well-defined quantities
(whatever approximation we use for the metric in the stellar
exterior). For example, for a HT model the circumferential stellar
radius on the equator is simply $\left[g_{\phi \phi}^{\rm
(HT)}(R_e,\theta=\pi/2)\right]^{1/2}$, where $R_e$ is the coordinate
radius. The stellar radius in Manko coordinates $\rho_e$ can be
evaluated using the procedure described in Section 4.1 of BS. Once
again, it's a simple matter to compute the corresponding equatorial
circumferential radius $\left[g_{\phi\phi}^{\rm
(Manko)}(\rho_e,z=0)\right]^{1/2}$ for the Manko models. Of course the
fact that we are using different approximations for the metric will
induce deviations between these stellar radii, but those deviations
are typically below 1 \% or so, even for the fastest rotating models.

In both panels of Fig.\ \ref{fig6}, lines from bottom to top give
$(1-|S|)$ as a function of the circumferential radius $R_{\rm circ}$
for different values of the rotation rate. Since the calculation only
applies to the vacuum exterior spacetime, each line starts from the
star's circumferential equatorial radius ($R_{\rm circ}\geq R_e$). The
plot shows that $(1-|S|)$ has the essential properties required of a
Non-Type-D-ness: first of all, slower rotating model are supposed to
be closer to Type D, and indeed $(1-|S|)$ is smaller; secondly, as
$R_{\rm circ}\to \infty$ the spacetime becomes asymptotically flat and
$(1-|S|)$ tends to zero.  

For slowly rotating models (roughly speaking, for $\epsilon<0.4$), the
lower lines in Fig.\ \ref{fig6} show that deviations from speciality
are at most of order 1 \%. So in most astrophysical situations it may
make sense to use a Type D approximation: that is, to consider the
exterior spacetime of a rotating star as a Kerr background, small
corrections being induced by the star's oblate structure. This Type D
approximation might be particularly useful for imposing boundary
conditions in the computation of gravitational waves from rotating
relativistic stars. Indeed, the gravitational wave amplitude dies out
only as $1/r$, while our analysis shows that effects due to deviations
from Type D die out faster than $1/r$. Therefore, far enough from the
star the gravitational wave amplitude should (in some sense) be large
enough that we can ignore the error made by using the Kerr
approximation. On the other hand, the Kerr metric usually has a
quadrupole moment that is 2 to 12 times smaller than that of a rapidly
rotating star (Laarakkers \& Poisson 1999). So, as long as we are in a
regime in which the quadrupole moment affects the gravitational wave
emission, the contribution of the quadrupole term will be underestimed
by the corresponding factor. Only when we are far enough away that
only the first-order terms in $\epsilon$ matter do the neutron star and
BH spacetimes agree. In this sense Kerr is not a good replacement for
HT to order $\epsilon^2$, because of the large difference in the
quadrupole moment. Using HT to order $\epsilon^2$ should give a much
more accurate waveform consistent to this order. A wave extraction
formalism which is consistent to second order in $\epsilon$ is therefore
worth developing.

\begin{figure} 
{
\centerline{\resizebox{9cm}{!}{\rotatebox{0}{\includegraphics{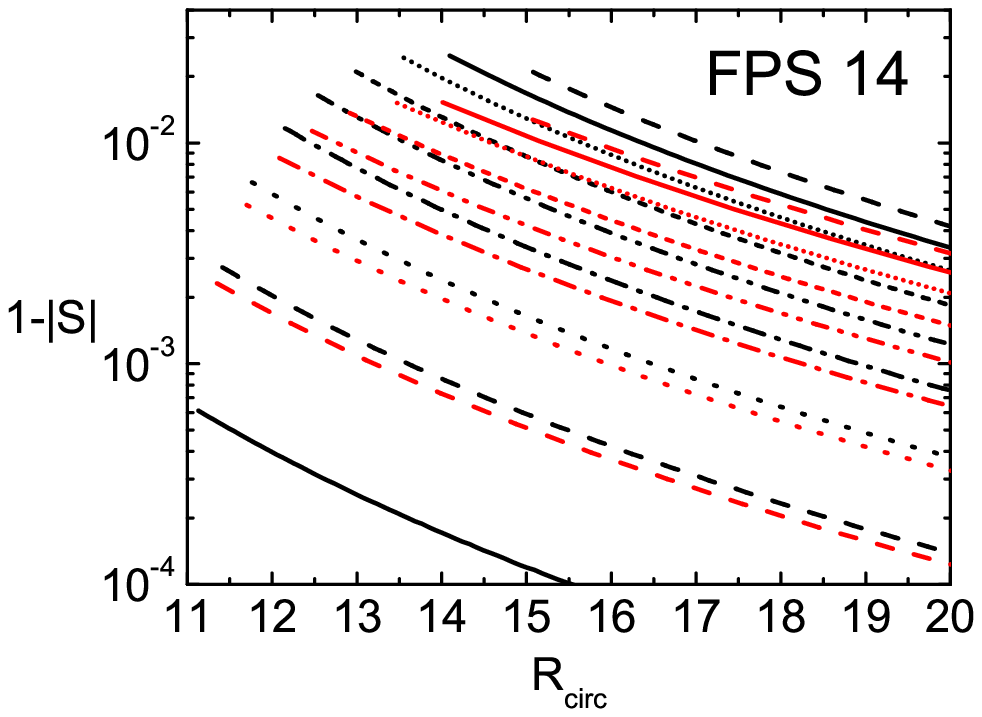}}}}
\centerline{\resizebox{9cm}{!}{\rotatebox{0}{\includegraphics{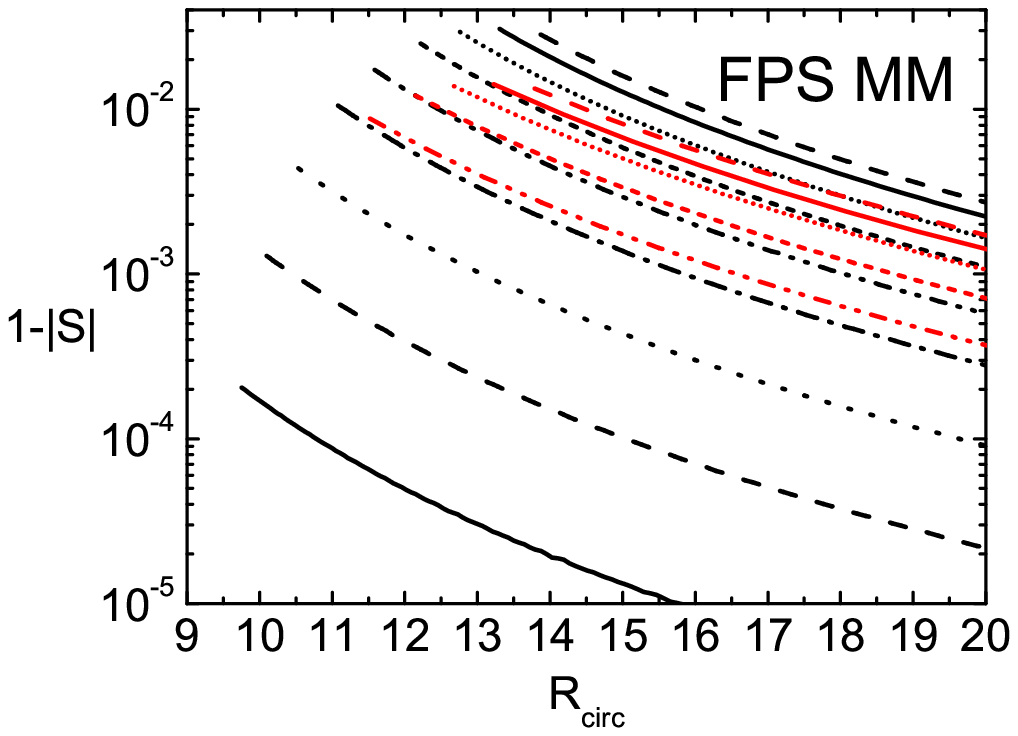}}}}
}
\caption{ 
Close-up view of Fig.\ \ref{fig6}. Once again, the top panel refers to
the FPS 14 sequence and the bottom panel to the FPS MM sequence. In
red we overplot the Non-Type-D-ness for the Manko matching models, to
visualize the difference with respect to the corresponding CST models.
}
\label{fig7}
\end{figure}

Fig.\ \ref{fig7} shows a close-up view of the strong field region of
Fig.\ \ref{fig6}. In red we overplot the Non-Type-D-ness we computed
for the matching Manko models (when they exist: recall that the Manko
solution can only be matched to fast rotating models). For both
sequences and for all values of the rotation rate the Manko metric is
closer to Type D than the corresponding numerical model. 

\begin{figure} 
{
\centerline{\resizebox{9cm}{!}{\rotatebox{0}{\includegraphics{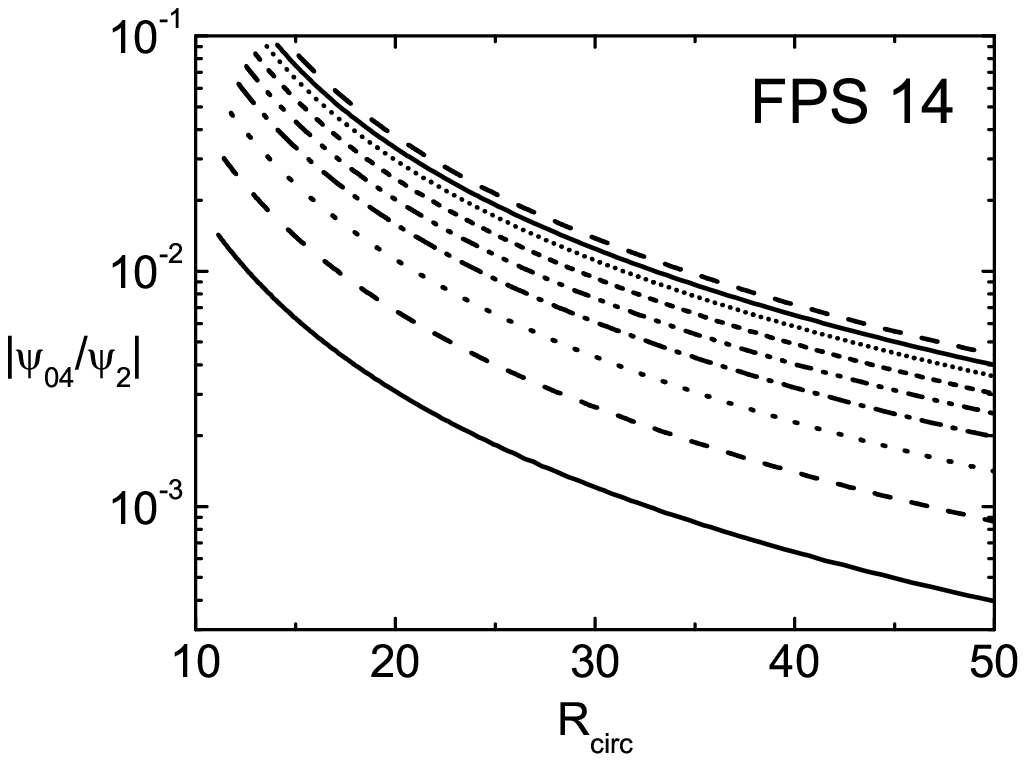}}}}
\centerline{\resizebox{9cm}{!}{\rotatebox{0}{\includegraphics{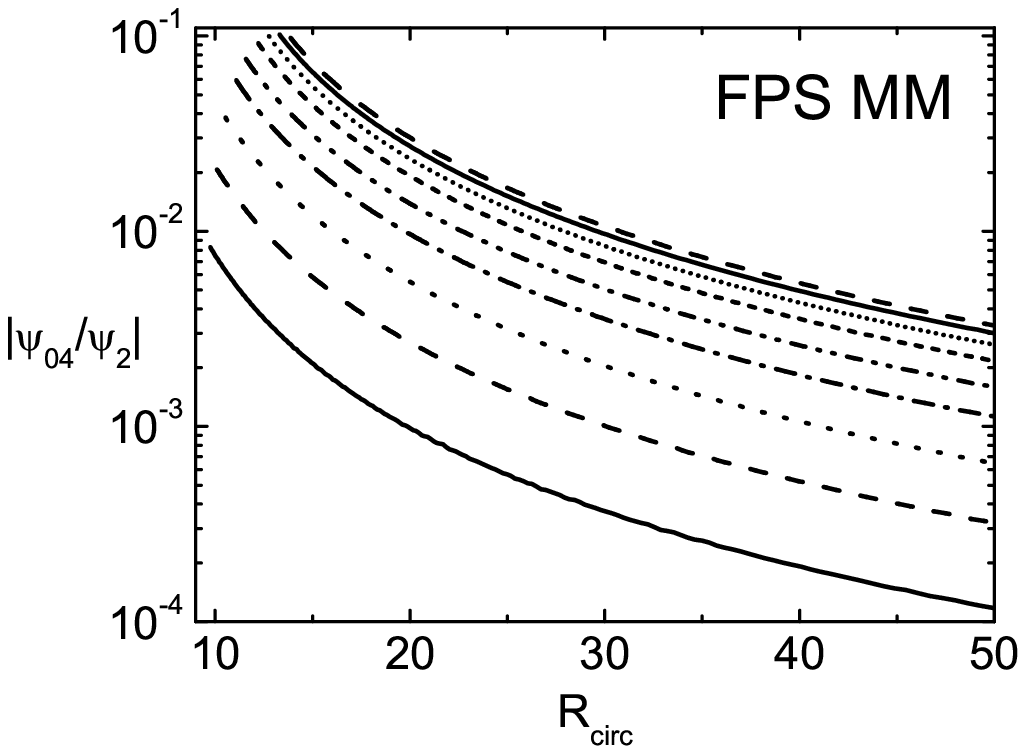}}}}
}
\caption{ 
Ratio $|\psi_{04}/\psi_2|$ for the FPS 14 (top panel) and FPS MM (bottom
panel) sequences, computed using the CST metric, as a function of the
circumferential radius $R_{\rm circ}$. See the caption of Fig.
\ref{fig6} for more details.
}
\label{fig5}
\end{figure}

In Fig. \ref{fig5} we plot the ratio $|\psi_{04}/\psi_2|$ as a
function of the circumferential radius $R_{\rm circ}$ for the CST
metric. As expected from our discussion in Sec. \ref{QK}, the
qualitative behavior is the same as in Fig. \ref{fig6}.
$|\psi_{04}/\psi_2|$ is not invariant ($\psi_{04}$ is a curvature
invariant, in the sense discussed above, but $\psi_2$ is not), but it
has the advantage of being dimensionless. One may want to develop a
`quasi-Type-D' approximation for perturbations of rotating NSs in the
framework of the NP formalism, similar to the Teukolsky formalism for
BHs. In this context the perturbed $\psi_0$ and $\psi_4$ would
represent radiation, typically decaying as $1/r$. Then the value of
the background $\psi_{04}$ would be a useful number to consider for
comparison. It is clear from Eq. (\ref{psi04HTasy}) for the HT
approximation and from Fig.\ \ref{fig5} for the CST metric that
$\psi_{04}$ decays faster than $1/r$.

\begin{figure} 
{
\centerline{\resizebox{9cm}{!}{\rotatebox{0}{\includegraphics{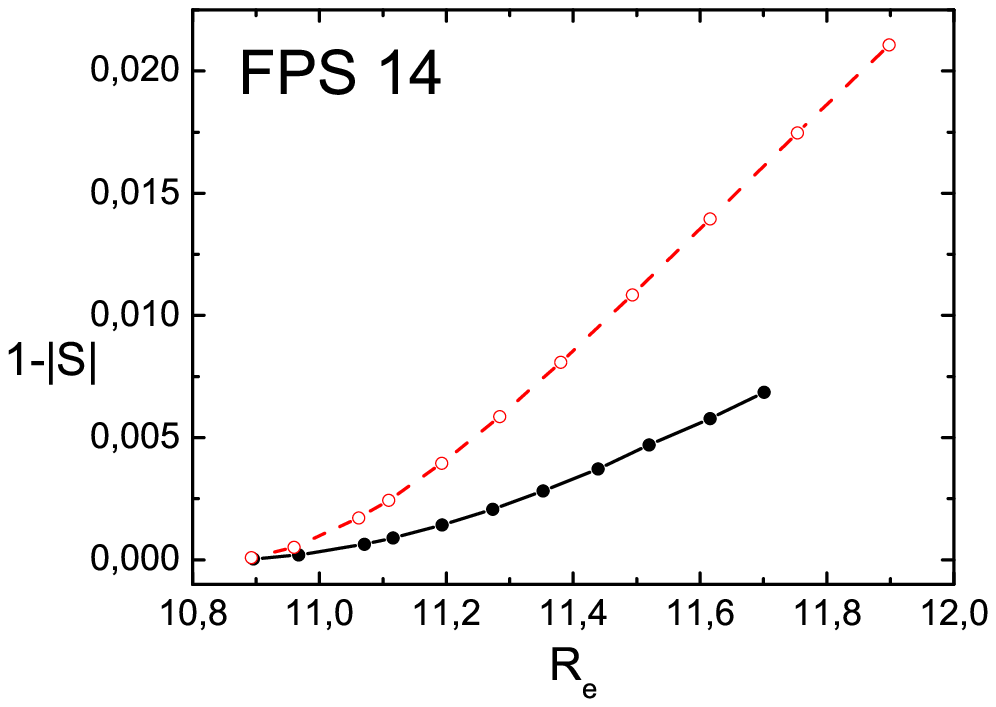}}}}
\centerline{\resizebox{9cm}{!}{\rotatebox{0}{\includegraphics{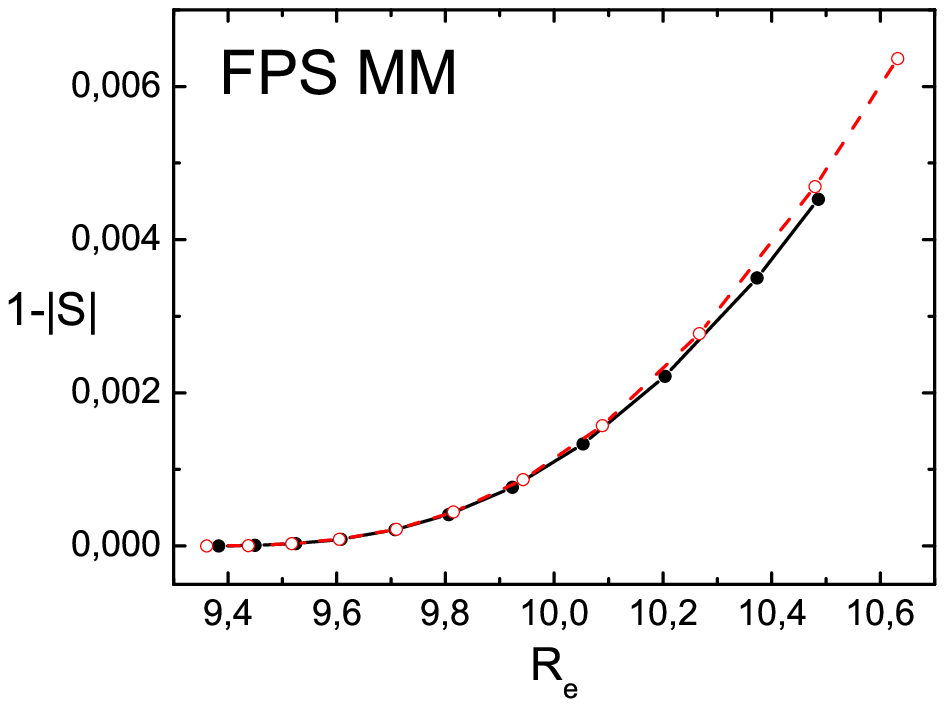}}}}
}
\caption{ 
Non-Type-D-ness evaluated at the star's circumferential equatorial
radius $R_e$, as a function of $R_e$, for the FPS 14 and FPS MM
sequences (top and bottom panel, respectively). Filled circles
correspond to CST models, empty circles correspond to matched HT
models.
}
\label{fig8}
\end{figure}

\begin{figure} 
{
\centerline{\resizebox{9cm}{!}{\rotatebox{0}{\includegraphics{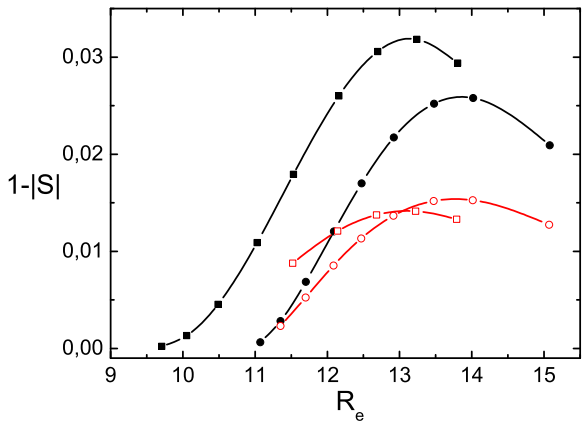}}}}
}
\caption{ 
Non-Type-D-ness evaluated at the star's circumferential equatorial
radius $R_e$, as a function of $R_e$, for the two sequences FPS 14
(circles) and FPS MM (squares). Filled symbols correspond to CST
models, empty symbols correspond to matching Manko models (when they
exist: cfr. BS).
}
\label{fig9}
\end{figure}

One might think that the value of the Non-Type-D-ness at the
equatorial stellar radius, which is a physically well-identified
location, could give an invariant measure of the deviation from
speciality of each rotating stellar model.  This is indeed the case
for slowly rotating stars. In Fig.\ \ref{fig8} we plot the value of
$(1-|S|)$ at the equatorial circumferential stellar radius $R_e$ as a
function of $R_e$ for our slowly rotating FPS 14 and FPS MM
sequences. This quantity has the behavior expected of a
Non-Type-D-ness in this rotational range: both for the CST metric and
for the HT metric it increases with $R_e$ (and with the rotation
rate). Furthermore, it is larger for the FPS 14 sequence than for the
FPS MM sequence. Once again, this is reasonable: low-mass models are
more oblate than high-mass models, and their exterior spacetime is
expected to deviate more from that of a BH.

From Fig.\ \ref{fig6} and Fig.\ \ref{fig7} it is clear that, at a
given circumferential radius, $(1-|S|)$ increases with the stellar
rotation rate.  However, for large rotation rates $(1-|S|)$ {\it at
the stellar equator} does {\it not} increase monotonically with
rotation: we show this peculiar effect in Fig.\ \ref{fig9}. In other
words, this single number cannot be used to measure the deviation from
speciality of a very fast rotating stellar model. To understand the
reason for this non-monotonic behavior we can look again at Fig.\
\ref{fig7}. At any given circumferential radius $R_{\rm circ}$,
$(1-|S|)$ does increase as the star spins faster. However the stellar
circumferential radius (the value of $R_{\rm circ}$ at which each line
starts) increases as well, due to the centrifugal deformation of the
star. The two effects compensate, so that $(1-|S|)$ at the stellar
radius shows a {\it maximum} as a function of the rotation rate.  From
Fig.\ \ref{fig9} we see that this maximum shows up at the same
rotation rate both for the CST metric and for the Manko matching
models.

\section{Conclusions}\label{conc}

In this paper we compared three different approaches to the
computation of rotating relativistic star spacetimes: the HT
slow-rotation approximation, the Manko analytic vacuum solution, and
numerical solutions of the full Einstein equations obtained using the
CST self-consistent field method.  We first integrated the HT
structure equations for five representative equations of state,
keeping terms up to second order in the slow-rotation parameter
$\epsilon$. Then we matched these models to the CST solutions,
imposing that the gravitational mass and angular momentum of the
models be the same. We estimated the limits of validity of the
slow-rotation expansion computing deviations in the spacetime's
quadrupole moments and in the ISCO radii at different rotation
rates. We found that deviations in the quadrupole moment are $\sim
20$~\% for pulsars spinning with a period $\simeq 1.5$~milliseconds
(the spin period of the fastest known pulsar, PSR J1939+2134).
However, for these same spin rates deviations in the ISCO radii are
always smaller than $\sim 1~\%$. Since the HT approximation gives
excellent predictions for ISCOs up to the fastest pulsar spin periods,
it can safely be used whenever a full numerical solution would be too
cumbersome to implement.

In the second part of the paper we focused on the exterior vacuum
spacetime. We compared the HT approximation and the Manko analytic
solution with numerical models using coordinate-independent
quantities. For all three spacetimes we first introduced symmetric
transverse frames, in which the only non-zero Weyl scalars are
$\psi_2$ and $\psi_0=\psi_4\equiv \psi_{04}$. From those frames we
then selected what we call the `quasi-Kinnersley' frame (cf. Nerozzi
{\it et al.} 2004): the frame in which $\psi_2$ is the {\it only}
non-zero scalar when we set the parameters to reduce the metric to
Type D. We computed $\psi_2$ and $\psi_{04}$ in this frame. The latter
is a scalar curvature invariant (Beetle \& Burko 2002) that vanishes
for Type D, and therefore its value is a measure of what one would
call `Non-Type-D-ness'. For the HT metric the use of the
quasi-Kinnersley frame led us to very simple analytic expressions for
$\psi_2$ and $\psi_{04}$. Due to their simplicity, our Eqs.\
(\ref{psi04HT}) and (\ref{psi2HT}) may be useful in different
contexts. Furthermore, from Eq. (\ref{psi04HT}) we reached the
important conclusion that, at leading order, the deviation from Type D
is linear in the deviation of the quadrupole moment from its Kerr
value, $(Q-Q_{\rm Kerr})$. Therefore neutron stars with smaller mass
and stiffer EOS deviate more from Type D.

Finally we evaluated the (scalar curvature invariant) speciality index
$S$ introduced by Baker \& Campanelli (2000). Being dimensionless,
$(1-|S|)$ is a better indicator of `Non-Type-D-ness'. As $\psi_{04}$,
it increases with rotation rate and goes to zero at infinity for any
given rotation rate. At leading order from Eq. (\ref{expS})
$(1-S)\propto (\psi_{04}/\psi_2)^2$. However, it is not easy to give a
single number characterizing the Non-Type-D-ness of a given
spacetime. One might think about using the value of $(1-|S|)$ {\it at
the stellar radius} for this purpose. This idea does not work in
practice: since stellar models become more and more oblate as rotation
increases, $(1-|S|)$ at the stellar radius eventually shows a {\it
maximum} as a function of the rotation rate.

Although deviations from Type D increase for fast rotation, our
results show that they can be expected to be rather small (less than a
few percent) for astrophysically relevant rotation rates. This
suggests that, in perturbative calculations of gravitational wave
emission, rotating star exteriors can reasonably be approximated as
Kerr-like spacetimes, or that one could develop an appropriate
quasi-Type D approximation. The tools we have illustrated here will
help to quantify the accuracy of such approximations, and hence to
estimate numerical errors involved in the calculations of
gravitational waves from rotating relativistic stars.

\begin{center}
\bf{Acknowledgements}
\end{center}

We would like to thank Lior Burko for very kindly providing us with
notes on the method to identify the three transverse frames, and the
anonymous referee for insightful comments. We also thank Marek
Abramowicz, Joachim Almergren, Nils Andersson, Christian Cherubini,
Wlodek Klu\'zniak, Kostas Kokkotas, John Miller, Andrea Nerozzi,
Luciano Rezzolla, Adam Stavridis, Nick Stergioulas, Arun Thampan and
Clifford Will for useful discussions. This work has been partly
supported by the EU Programme `Improving the Human Research Potential
and the Socio-Economic Knowledge Base' (Research Training Network
Contract HPRN-CT-2000-00137). FW is supported by a EPSRC studentship.

\clearpage

\begin{table}
\centering 
\caption{ 
Results of the HT integrations for EOS A. From left to right we list:
the central energy density $\rho_c$; the stellar radius $R$,
gravitational mass $M$ and binding energy $E_B$ for the non-rotating
configuration; $\Omega^*=(M/R^3)^{1/2}$. Then we give quantities of
order $\epsilon$ in rotation: the radius of gyration $R_g^*$ and the
frame dragging function $\omega_c^*$ at the center (normalized by
$\Omega^*$). In the last five columns we give quantities of order
$\epsilon^2$, namely the rotational corrections on radius ($\delta
R^*/R$), mass ($\delta M^*/M$) and binding energy ($\delta
E_B^*/E_B$), the quadrupole moment $Q^*/(MR^2)$ and the eccentricity
$e_s^*$.  Units are given where appropriate.
} \vskip 12pt
\renewcommand{\arraystretch}{1}
\begin{tabular}{@{}|c|cccc|cc|ccccc|@{}}
\hline
& $\sim\epsilon^0$ & & & &$\sim\epsilon$ & &$\sim \epsilon^2$ & & & &\\
$\rho_c$ (g~cm$^{-3}$) &$R$ (km) &$M/M_\odot$ &$E_B/M_\odot$ &$\Omega^*$ (km$^{-1}$) &$R_g^*/R$ &$\omega_c^*/\Omega^*$
&$\delta R^*/R$ &$\delta M^*/M$ &$\delta E_B^*/E_B$ &$Q^*/(MR^2)$ &$e_s^*$
\\
\hline
1.25E15 &9.90 &1.04 &0.0933&0.0398 &0.583 &0.602 &0.120 &0.338  &0.417 &0.115    &1.13 \\
1.35E15 &9.86 &1.12 &0.111 &0.0416 &0.592 &0.571 &0.114 &0.330  &0.395 &0.115    &1.12 \\
1.46E15 &9.81 &1.20 &0.129 &0.0434 &0.601 &0.540 &0.107 &0.322  &0.372 &0.113    &1.11 \\
1.58E15 &9.74 &1.27 &0.148 &0.0451 &0.609 &0.509 &0.100 &0.312  &0.347 &0.112    &1.11 \\
1.72E15 &9.66 &1.34 &0.168 &0.0468 &0.617 &0.479 &0.0936 &0.302 &0.321 &0.110  &1.10  \\
1.86E15 &9.57 &1.40 &0.187 &0.0486 &0.624 &0.450 &0.0870 &0.291 &0.295 &0.108  &1.08  \\
2.01E15 &9.47 &1.45 &0.205 &0.0503 &0.631 &0.421 &0.0805 &0.280 &0.268 &0.107  &1.07  \\
2.18E15 &9.36 &1.50 &0.223 &0.0520 &0.638 &0.394 &0.0742 &0.270 &0.242 &0.105  &1.06  \\
2.36E15 &9.25 &1.54 &0.238 &0.0537 &0.644 &0.368 &0.0683 &0.260 &0.216 &0.103  &1.05  \\
2.56E15 &9.13 &1.58 &0.252 &0.0553 &0.649 &0.344 &0.0628 &0.251 &0.191 &0.102  &1.04  \\
2.77E15 &9.00 &1.60 &0.264 &0.0570 &0.654 &0.321 &0.0576 &0.242 &0.167 &0.101  &1.03  \\
3.00E15 &8.87 &1.62 &0.274 &0.0586 &0.659 &0.300 &0.0528 &0.233 &0.144 &0.100  &1.02  \\
3.25E15 &8.75 &1.64 &0.281 &0.0601 &0.663 &0.280 &0.0485 &0.225 &0.122 &0.0995 &1.01  \\
3.52E15 &8.62 &1.65 &0.286 &0.0617 &0.667 &0.261 &0.0445 &0.218 &0.100 &0.0988 &1.00  \\
3.81E15 &8.49 &1.65 &0.289 &0.0632 &0.670 &0.244 &0.0408 &0.211 &0.0797&0.0982 &0.995 \\
4.13E15 &8.36 &1.66 &0.290 &0.0647 &0.673 &0.228 &0.0376 &0.205 &0.0601&0.0977 &0.987 \\
\hline
\end{tabular}
\label{A}
\end{table}

\begin{table}
\centering 
\caption{ 
Same as in Table \ref{A}, but for EOS AU.
} \vskip 12pt
\renewcommand{\arraystretch}{1}
\begin{tabular}{@{}|c|cccc|cc|ccccc|@{}}
\hline
& $\sim\epsilon^0$ & & & &$\sim\epsilon$ & &$\sim \epsilon^2$ & & & &\\
$\rho_c$ (g~cm$^{-3}$) &$R$ (km) &$M/M_\odot$ &$E_B/M_\odot$ &$\Omega^*$ (km$^{-1}$) &$R_g^*/R$ &$\omega_c^*/\Omega^*$
&$\delta R^*/R$ &$\delta M^*/M$ &$\delta E_B^*/E_B$ &$Q^*/(MR^2)$ &$e_s^*$
\\
\hline
7.62E14 &10.3 &0.602&0.0254&0.0285 &0.513 &0.763 &0.150 &0.322  &0.461   &0.0946  &1.12  \\
8.36E14 &10.3 &0.742&0.0413&0.0317 &0.539 &0.718 &0.142 &0.342  &0.468   &0.107   &1.13  \\
9.16E14 &10.3 &0.895&0.0632&0.0347 &0.562 &0.670 &0.134 &0.351  &0.460   &0.115   &1.14  \\
1.00E15 &10.3 &1.05 &0.0914&0.0375 &0.582 &0.620 &0.124 &0.350  &0.439   &0.119   &1.14  \\
1.10E15 &10.4 &1.22 &0.128 &0.0402 &0.601 &0.566 &0.112 &0.342  &0.406   &0.120   &1.13  \\
1.21E15 &10.4 &1.40 &0.175 &0.0429 &0.621 &0.508 &0.0989 &0.327 &0.361   &0.120  &1.12  \\
1.32E15 &10.4 &1.57 &0.229 &0.0456 &0.639 &0.450 &0.0850 &0.310 &0.311   &0.119  &1.10  \\
1.45E15 &10.3 &1.72 &0.285 &0.0480 &0.656 &0.397 &0.0716 &0.291 &0.260   &0.118  &1.08  \\
1.59E15 &10.3 &1.84 &0.338 &0.0502 &0.671 &0.349 &0.0594 &0.273 &0.210   &0.118  &1.06  \\
1.74E15 &10.2 &1.94 &0.383 &0.0522 &0.684 &0.309 &0.0489 &0.258 &0.165   &0.118  &1.04  \\
1.91E15 &10.1 &2.01 &0.421 &0.0540 &0.694 &0.274 &0.0399 &0.244 &0.124   &0.118  &1.03  \\
2.09E15 &9.94 &2.06 &0.451 &0.0556 &0.703 &0.244 &0.0323 &0.232 &0.0873  &0.119  &1.01  \\
2.29E15 &9.81 &2.10 &0.473 &0.0573 &0.711 &0.217 &0.0256 &0.222 &0.0533  &0.120  &1.00  \\
2.51E15 &9.68 &2.12 &0.488 &0.0588 &0.717 &0.194 &0.0199 &0.213 &0.0226  &0.122  &0.990 \\
2.76E15 &9.54 &2.13 &0.497 &0.0602 &0.722 &0.174 &0.0150 &0.205 &-0.00527&0.123  &0.980 \\
3.02E15 &9.40 &2.13 &0.499 &0.0616 &0.727 &0.156 &0.0110 &0.198 &-0.0305 &0.123  &0.971 \\
\hline
\end{tabular}
\label{AU}
\end{table}

\clearpage

\begin{table}
\centering 
\caption{ 
Same as in Table \ref{A}, but for EOS FPS. 
} \vskip 12pt
\renewcommand{\arraystretch}{1}
\begin{tabular}{@{}|c|cccc|cc|ccccc|@{}}
\hline
& $\sim\epsilon^0$ & & & &$\sim\epsilon$ & &$\sim \epsilon^2$ & & & &\\
$\rho_c$ (g~cm$^{-3}$) &$R$ (km) &$M/M_\odot$ &$E_B/M_\odot$ &$\Omega^*$ (km$^{-1}$) &$R_g^*/R$ &$\omega_c^*/\Omega^*$
&$\delta R^*/R$ &$\delta M^*/M$ &$\delta E_B^*/E_B$ &$Q^*/(MR^2)$ &$e_s^*$
\\
\hline
8.03E14 &11.2 &0.873&0.0533&0.0303 &0.546 &0.695 &0.139 &0.339  &0.454 &0.108   &1.13  \\
8.84E14 &11.1 &0.986&0.0704&0.0324 &0.560 &0.659 &0.132 &0.339  &0.441 &0.111   &1.13  \\
9.73E14 &11.1 &1.10 &0.0901&0.0345 &0.573 &0.621 &0.125 &0.335  &0.422 &0.112   &1.13  \\
1.07E15 &11.0 &1.21 &0.112 &0.0364 &0.585 &0.584 &0.117 &0.328  &0.399 &0.112   &1.12  \\
1.18E15 &10.9 &1.31 &0.135 &0.0384 &0.595 &0.547 &0.109 &0.318  &0.372 &0.111   &1.11  \\
1.30E15 &10.8 &1.40 &0.159 &0.0403 &0.605 &0.512 &0.101 &0.308  &0.344 &0.109   &1.10  \\
1.43E15 &10.7 &1.48 &0.183 &0.0421 &0.614 &0.477 &0.0935 &0.296 &0.315 &0.107  &1.09  \\
1.57E15 &10.6 &1.56 &0.206 &0.0439 &0.622 &0.444 &0.0861 &0.285 &0.286 &0.105  &1.08  \\
1.73E15 &10.5 &1.62 &0.227 &0.0457 &0.630 &0.412 &0.0790 &0.273 &0.256 &0.103  &1.07  \\
1.91E15 &10.3 &1.67 &0.247 &0.0475 &0.637 &0.382 &0.0721 &0.261 &0.227 &0.102  &1.05  \\
2.10E15 &10.1 &1.71 &0.264 &0.0493 &0.643 &0.353 &0.0657 &0.250 &0.198 &0.0999 &1.04  \\
2.31E15 &9.97 &1.75 &0.278 &0.0510 &0.649 &0.327 &0.0598 &0.240 &0.170 &0.0985 &1.03  \\
2.54E15 &9.80 &1.77 &0.289 &0.0528 &0.654 &0.302 &0.0543 &0.230 &0.144 &0.0974 &1.02  \\
2.80E15 &9.62 &1.79 &0.297 &0.0545 &0.658 &0.279 &0.0495 &0.221 &0.118 &0.0964 &1.01  \\
3.08E15 &9.45 &1.80 &0.301 &0.0561 &0.662 &0.258 &0.0451 &0.213 &0.0941&0.0955 &0.999 \\
3.39E15 &9.28 &1.80 &0.303 &0.0577 &0.666 &0.239 &0.0413 &0.206 &0.0713&0.0947 &0.990 \\
\hline
\end{tabular}
\label{FPS}
\end{table}

\begin{table}
\centering 
\caption{ 
Same as in Table \ref{A}, but for EOS L. 
} \vskip 12pt
\renewcommand{\arraystretch}{1}
\begin{tabular}{@{}|c|cccc|cc|ccccc|@{}}
\hline
& $\sim\epsilon^0$ & & & &$\sim\epsilon$ & &$\sim \epsilon^2$ & & & &\\
$\rho_c$ (g~cm$^{-3}$) &$R$ (km) &$M/M_\odot$ &$E_B/M_\odot$ &$\Omega^*$ (km$^{-1}$) &$R_g^*/R$ &$\omega_c^*/\Omega^*$
&$\delta R^*/R$ &$\delta M^*/M$ &$\delta E_B^*/E_B$ &$Q^*/(MR^2)$ &$e_s^*$
\\
\hline
2.35E14 &14.9 &0.362&0.00513&0.0127 &0.426  &0.886 &0.173 &0.234  &0.396&0.0522  &1.07 \\
2.65E14 &14.5 &0.502&0.0119 &0.0156 &0.473  &0.853 &0.164 &0.302  &0.459&0.0804  &1.11 \\
3.00E14 &14.5 &0.680&0.0242 &0.0182 &0.510  &0.812 &0.159 &0.350  &0.500&0.105   &1.14 \\
3.39E14 &14.6 &0.896&0.0445 &0.0206 &0.540  &0.765 &0.153 &0.377  &0.517&0.122   &1.16 \\
3.83E14 &14.8 &1.14 &0.0751 &0.0228 &0.564  &0.711 &0.144 &0.387  &0.510&0.132   &1.17 \\
4.33E14 &15.0 &1.40 &0.116  &0.0249 &0.585  &0.655 &0.134 &0.383  &0.485&0.135   &1.16 \\
4.89E14 &15.1 &1.67 &0.168  &0.0268 &0.603  &0.597 &0.121 &0.370  &0.448&0.133   &1.15 \\
5.53E14 &15.1 &1.91 &0.229  &0.0286 &0.619  &0.540 &0.108 &0.351  &0.402&0.130   &1.14 \\
6.24E14 &15.1 &2.14 &0.292  &0.0303 &0.633  &0.486 &0.0955 &0.330 &0.353&0.126  &1.12 \\
7.06E14 &15.0 &2.32 &0.353  &0.0319 &0.645  &0.437 &0.0834 &0.309 &0.304&0.122  &1.10 \\
7.97E14 &14.8 &2.46 &0.407  &0.0334 &0.656  &0.393 &0.0727 &0.290 &0.259&0.118  &1.08 \\
9.01E14 &14.6 &2.56 &0.450  &0.0348 &0.664  &0.355 &0.0636 &0.273 &0.218&0.116  &1.06 \\
1.02E15 &14.4 &2.64 &0.483  &0.0361 &0.671  &0.322 &0.0560 &0.258 &0.182&0.113  &1.05 \\
1.15E15 &14.2 &2.68 &0.504  &0.0373 &0.676  &0.294 &0.0498 &0.246 &0.151&0.111  &1.03 \\
1.30E15 &13.9 &2.70 &0.514  &0.0383 &0.679  &0.271 &0.0449 &0.235 &0.125&0.110  &1.02 \\
1.47E15 &13.7 &2.70 &0.517  &0.0393 &0.682  &0.251 &0.0411 &0.226 &0.102&0.108  &1.01 \\
\hline
\end{tabular}
\label{L}
\end{table}

\clearpage

\begin{table}
\centering 
\caption{ 
Same as in Table \ref{A}, but for EOS APRb. 
} \vskip 12pt
\renewcommand{\arraystretch}{1}
\begin{tabular}{@{}|c|cccc|cc|ccccc|@{}}
\hline
& $\sim\epsilon^0$ & & & &$\sim\epsilon$ & &$\sim \epsilon^2$ & & & &\\
$\rho_c$ (g~cm$^{-3}$) &$R$ (km) &$M/M_\odot$ &$E_B/M_\odot$ &$\Omega^*$ (km$^{-1}$) &$R_g^*/R$ &$\omega_c^*/\Omega^*$
&$\delta R^*/R$ &$\delta M^*/M$ &$\delta E_B^*/E_B$ &$Q^*/(MR^2)$ &$e_s^*$
\\
\hline
6.16E14 &12.1 &0.654 &0.0243&0.0234 &0.497 &0.771 &0.152 &0.294   &0.430   &0.0835 &1.10  \\
6.78E14 &11.9 &0.776 &0.0368&0.0261 &0.518 &0.733 &0.145 &0.311   &0.438   &0.0927 &1.11  \\
7.46E14 &11.8 &0.917 &0.0546&0.0288 &0.539 &0.691 &0.137 &0.322   &0.436   &0.100  &1.12  \\
8.21E14 &11.7 &1.07  &0.0787&0.0315 &0.559 &0.644 &0.128 &0.326  &0.423   &0.105  &1.12 \\
9.04E14 &11.6 &1.24  &0.110 &0.0341 &0.578 &0.594 &0.118 &0.324  &0.400   &0.108  &1.12 \\
9.95E14 &11.6 &1.40  &0.147 &0.0366 &0.596 &0.543 &0.107 &0.316  &0.369   &0.109  &1.11 \\
1.10E15 &11.5 &1.56  &0.190 &0.0390 &0.613 &0.492 &0.0953 &0.305 &0.331   &0.109 &1.10 \\
1.21E15 &11.4 &1.70  &0.236 &0.0413 &0.629 &0.443 &0.0837 &0.291 &0.289   &0.109 &1.08 \\
1.33E15 &11.3 &1.83  &0.282 &0.0435 &0.643 &0.397 &0.0724 &0.276 &0.247   &0.108 &1.07 \\
1.46E15 &11.1 &1.94  &0.325 &0.0456 &0.656 &0.355 &0.0619 &0.262 &0.204   &0.108 &1.05 \\
1.61E15 &11.0 &2.02  &0.365 &0.0475 &0.668 &0.316 &0.0522 &0.248 &0.162   &0.108 &1.03 \\
1.77E15 &10.8 &2.09  &0.398 &0.0494 &0.678 &0.281 &0.0435 &0.235 &0.123   &0.108 &1.02 \\
1.95E15 &10.6 &2.14  &0.425 &0.0512 &0.688 &0.250 &0.0359 &0.224 &0.0865  &0.109 &1.01 \\
2.15E15 &10.5 &2.17  &0.444 &0.0529 &0.695 &0.223 &0.0292 &0.214 &0.0525  &0.110 &0.993\\
2.36E15 &10.3 &2.19  &0.457 &0.0545 &0.702 &0.199 &0.0235 &0.205 &0.0212  &0.111 &0.981\\
2.60E15 &10.1 &2.20  &0.463 &0.0560 &0.707 &0.178 &0.0187 &0.198 &-0.00756&0.112 &0.971\\
\hline
\end{tabular}
\label{APRb}
\end{table}

\clearpage


\begin{table}
\centering 
\caption{ 
Properties of HT models matching the slowest rotating CST models
listed in Tables 1-5 of BS.  For each EOS, on top we present matching
models for the $M=1.4 M_\odot$ (`14') sequence; separated by an
horizontal line, we give matching models for the maximum-mass (`MM')
sequence.  From left to right we list: the central energy density
$\bar \rho_c$ for which we find a matching solution (if such a central
energy density exists); the value of $\epsilon=\bar J/J^*$ at the given
central energy density; the gravitational mass $\bar M$ and the angular
momentum $\bar J$, in geometrical units; the value of the quadrupole
moment predicted by the HT equations of structure, $Q^{\rm HT}$, in
geometrical units; the percentage deviation of the HT quadrupole
moment from its value $Q$ for the CST spacetime.
} \vskip 12pt
\renewcommand{\arraystretch}{1}
\begin{tabular}{@{}|lcccccc|@{}}
\hline
EOS &$\bar \rho_c$ (g~cm$^{-3}$) &$\epsilon$ &$\bar M$ (km) &$\bar J$ (km$^2$) &$Q^{\rm HT}$ (km$^3$) & $\delta Q$ $(\%)$ \\
\hline
\hline
  A 14  &1.8110E15 &0.23236 &2.0729 &0.8121  &1.1053& 10.4\\
        &1.7238E15 &0.40060 &2.0800 &1.3267  &3.2644& 22.9\\
        &1.6152E15 &0.55642 &2.0870 &1.7037  &6.2084& 41.8\\
        &1.4476E15 &0.76898 &2.0940 &2.0220  &11.372& 84.2\\
\hline	   
  A MM  &3.4253E15 &0.19954 &2.4519 &0.9857  &0.7192& 15.4\\
        &3.0369E15 &0.32653 &2.4608 &1.5731  &2.0107&  21.3\\
        &2.6524E15 &0.46715 &2.4726 &2.1471  &4.2850&  32.4\\
        &2.2177E15 &0.65300 &2.4870 &2.7173  &8.6795&  59.7\\
\hline	   
\hline	   
 AU 14  &1.1916E15 &0.24345 &2.0734 &0.8704  &1.5633&  10.4\\
        &1.1666E15 &0.40245 &2.0794 &1.3551  &4.1490&  23.3\\
        &1.1217E15 &0.59998 &2.0868 &1.7984  &8.6917&  49.4\\
        &1.0426E15 &0.87088 &2.0925 &2.0808  &16.160&  109.8\\
\hline	   
 AU MM  &2.5557E15 &0.21819 &3.1645 &1.9389  &1.6913&  23.5\\
        &2.2777E15 &0.36512 &3.1838 &3.1382  &4.7843&  29.3\\
        &2.0247E15 &0.51420 &3.2068 &4.1755  &9.4664&  40.5\\
        &1.7585E15 &0.69420 &3.2302 &5.0587  &16.844&  66.1\\
\hline	   
\hline	   
FPS 14  &1.2642E15 &0.25144 &2.0720 &0.8722  &1.6662&  10.2\\
        &1.2195E15 &0.39258 &2.0769 &1.2971  &4.0094&  21.1\\
        &1.1565E15 &0.54440 &2.0825 &1.6649  &7.5270&  39.0\\
        &1.0626E15 &0.74208 &2.0881 &1.9766  &13.279&  74.8\\
\hline	   
FPS MM  &2.8003E15 &0.20109 &2.6648 &1.1604  &0.9524&  14.7\\
        &2.4765E15 &0.32814 &2.6740 &1.8467  &2.6577&  20.8\\
        &2.1574E15 &0.46924 &2.6864 &2.5193  &5.6832&  32.1\\
        &1.7968E15 &0.65491 &2.7014 &3.1853  &11.536&  59.7\\
\hline	   
\hline	   
  L 14  &4.2540E14 &0.25862 &2.0717 &0.9726  &4.0544&  10.4\\
        &4.1435E14 &0.43170 &2.0760 &1.5007  &10.756&  27.1\\
        &3.9803E14 &0.61954 &2.0802 &1.8991  &20.412&  54.6\\
\hline	   
  L MM  &1.2480E15 &0.18716 &4.0130 &2.5589  &3.0214&  13.8\\
        &1.0717E15 &0.32762 &4.0297 &4.3607  &9.6976&  21.0\\
        &9.2990E14 &0.47634 &4.0515 &6.0142  &21.145&  34.0\\
        &7.8669E14 &0.66069 &4.0754 &7.4955  &41.295&  63.6\\
\hline	   
\hline	   
APRb 14 &9.7905E14 &0.27032 &2.0753 &0.9349  &2.1641&  10.1\\
        &9.6661E14 &0.35913 &2.0781 &1.2034  &3.7615&  17.4\\
        &9.5231E14 &0.44237 &2.0810 &1.4276  &5.6017&  25.5\\
        &9.3547E14 &0.52688 &2.0838 &1.6238  &7.7629&  35.6\\
        &9.1459E14 &0.62069 &2.0866 &1.8017  &10.442&  49.1\\
\hline	   
APRb MM &2.4553E15 &0.17324 &3.2640 &1.6051  &1.1343&  25.3\\
        &2.3001E15 &0.24355 &3.2705 &2.2296  &2.2707&  23.0\\
        &2.0765E15 &0.36617 &3.2864 &3.2541  &5.2083&  28.3\\
        &1.8261E15 &0.52120 &3.3086 &4.3615  &10.632&  39.8\\
        &1.6702E15 &0.63017 &3.3232 &4.9670  &15.494&  54.0\\
        &1.4393E15 &0.81746 &3.3393 &5.5922  &25.448&  94.7\\
\hline
\hline
\end{tabular}
\label{Matching}
\end{table}

\end{document}